\documentclass[prd,preprint,eqsecnum,nofootinbib,amsmath,amssymb,
               tightenlines,floatfix]{revtex4}
\input epsf
\def\be{\begin{equation}}
\def\ee{\end{equation}}
\def\bea{\begin{eqnarray}}
\def\eea{\end{eqnarray}}
\def\centerbox#1#2{\centerline{\epsfxsize=#1\textwidth\epsfbox{#2}}}
\def\alphas{\alpha_{\rm s}}
\def\fbar{\bar{f}_1}
\def\fbarr{\bar{f}_2}
\def\tchi{{\tilde{\chi}}}
\def\bchi{{\bar{\chi}}}
\def\dhat#1#2{{\hat{#1}_{\langle r} \hat{#2}_{s \rangle}}}
\def\symindex#1#2#3#4{{#1_{\langle #3}#2_{#4 \rangle}}}
\def\Mrow{Mr\'owczy\'nski}
\def\nc{N_{\rm c}}
\def\nf{n_{\rm f}}
\def\h{{\bf h}}
\def\q{{\bf q}}
\def\p{{\bf p}}
\def\v{{\bf v}}
\def\k{{\bf k}}
\def\x{{\bf x}}
\def\F{{\bf F}}
\def\mD{m_{\rm D}}
\def\thelog{\ln \frac{\omega{+}q}{\omega{-}q}\:}
\def\cf{C_{\rm f}}
\def\df{d_{\rm f}}
\def\ca{C_{\rm A}}
\def\da{d_{\rm A}}
\def\Eq#1{Eq.~(\ref{#1})}
\def\M{{\cal M}}
\def\O{{\cal O}}
\def\C{{\cal C}}

\def\Tr{\:{\rm Tr}\:}
\def\Re{\:{\rm Re}\:}

\def\dsig{\delta \Sigma}
\def\ppm{{\pm}}
\def\commut#1#2{\Big[#1\,,\, #2\Big]}
\def\Baieretal{Baier {\it et al}}
\def\Ansatz{{\it Ansatz}}

\def\nott#1{\setbox0=\hbox{$#1$}                
   \dimen0=\wd0                                 
   \setbox1=\hbox{/} \dimen1=\wd1               
   \ifdim\dimen0>\dimen1                        
      \rlap{\hbox to \dimen0{\hfil/\hfil}}      
      #1                                        
   \else                                        
      \rlap{\hbox to \dimen1{\hfil$#1$\hfil}}   
      /                                         
   \fi}                                         %

\def\lsim{\mbox{~{\raisebox{0.4ex}{$<$}}\hspace{-1.1em}
        {\raisebox{-0.6ex}{$\sim$}}~}}

\begin{document}

\title{ Second order hydrodynamic coefficients from kinetic theory }

\author{Mark Abraao York and Guy D.\ Moore} 
\affiliation{McGill University Dept.\ of Physics, 3600 rue University,
    Montr\'eal QC H3A 2T8 Canada}

\date{October 2008: v2 February 2009}

\begin{abstract}
{In a relativistic setting, hydrodynamic calculations 
 which include shear viscosity (which is first order in an expansion in
 gradients of the flow velocity) are unstable and acausal unless they also
 include terms to second order in
 gradients.  To date such terms have only been computed in
 supersymmetric ${\cal N}{=}4$ Super-Yang-Mills theory at infinite coupling.
 Here we compute these second-order hydrodynamic coefficients in
 weakly coupled QCD, perturbatively to leading order in the QCD coupling,
 using kinetic theory.  We also compute them in QED and scalar $\lambda\phi^4$
 theory.
}
\end{abstract}

\maketitle
\thispagestyle{empty}

\section{Introduction and results}
\label{sec:intro}

Recently the Relativistic Heavy Ion Collider (RHIC) at Brookhaven has
successfully created the quark-gluon plasma.  Measurements of elliptic
flow \cite{RHIC_xpt} indicate collective fluid behavior which implies a
startlingly low viscosity \cite{sQGP}.  Actually, measured in Poise the
viscosity is enormously large; but this is expected of such a hot and
dense system.  It has recently been argued \cite{KSS} that viscosity
naturally scales with entropy density.  Their ratio $\eta/s$ is
dimensionless [in natural units, used throughout; restoring $\hbar$ and
  $c$, it has units of $\hbar$] and is conjectured to be bounded below by
$\eta/s \geq 1/4\pi$ (see however \cite{bound_violated}).

It is believed that the quark-gluon plasma created at RHIC displays a
viscosity relatively close to this bound.  But it is important to
quantify this by comparing experimental results for elliptic flow
spectra to the predictions of viscous hydrodynamics simulations.
Several groups are engaged in this
\cite{Romatschke,Heinz,Teaney,Chaudhuri,Molnar},
but it is not as simple as adding a viscosity term to the ideal
hydrodynamical equations.  Indeed, it has been known for decades
that relativistic Navier-Stokes equations are acausal and unstable
\cite{Muller,IsraelStewart,Hiscock}.%
\footnote%
    {The easy way to understand this is to note that Navier-Stokes
      equations are Euler equations plus a momentum-diffusion term, with
      the viscosity as the momentum-diffusion coefficient.  But
      diffusion equations possess infinite propagation speeds for
      information, which is problematic in a relativistic setting.}

Viscosity is just the first-order term in a gradient-expansion of
corrections to ideal Eulerian hydrodynamics; Israel and Stewart showed 30
years ago that the stability problems could be repaired by the inclusion
of certain second-order terms as well \cite{IsraelStewart}.  This is the
guiding philosophy for most recent viscous hydrodynamics studies of the
quark-gluon plasma.

However, once one allows for some second-order in gradients terms, it
seems wise to at least consider all second-order terms which could
appear and to make an estimate of their size relative to the shear
viscosity.  This program was begun recently by \Baieretal\
\cite{Baieretal}, who
showed that, with the additional assumption of conformal invariance (a
good approximation in QCD if the temperature is well above the QCD
transition/crossover temperature of $\sim 170$ MeV), there are five
second-order coefficients, one of which is only relevant in curved
space.

It would be valuable to have a reasonable estimate of the size of these
second-order coefficients, or an estimate of how they scale with the
shear viscosity.  \Baieretal\ and the Tata group \cite{Tata} have
given one estimate, by evaluating the five coefficients in a toy model
for QCD, strongly coupled ${\cal N}{=}4$ Super-Yang-Mills theory
(see also \cite{Natsuume}).  Here
we evaluate the five second-order coefficients in QCD to leading order
in the weak coupling expansion, using kinetic theory.  In the
thermal field theory setting the coupling expansion is not believed to
converge very well (see for instance \cite{CaronMoore}), so weakly
coupled QCD should also be
viewed as a ``toy model'' for QCD at realistic couplings.  However we
hope that the combined insight from the two ``toy models'' give a
reasonable idea of the expected scaling of these second-order
coefficients relative to shear viscosity.

We evaluate the flat-space coefficients in Section \ref{sec:kinetic} and the
curved-space coefficient in Section \ref{sec:Kubo}.
We then give an extensive discussion, in Section \ref{sec:discussion}, of the
physical interpretation of each
second-order transport coefficient, and some interesting physical issues which
arise in their computation.  Certain technical
details involving nonlinear corrections arising through plasma screening are
postponed to Appendix \ref{sec:appendix}.
%
%
But we will finish introducing the problem and present
the main results and conclusions here.

All hydrodynamic approaches are based on stress-energy conservation,
\be
\partial_\mu T^{\mu\nu} = 0 \,,
\ee
which is 4 equations for 10 unknowns.%
\footnote{
    Here we only consider systems with vanishing densities of other
    conserved charges such as baryon number.
    }
The other 6 equations are
established by gradient expanding the form of $T^{\mu\nu}$ about its
equilibrium form.  In the absence of nonzero conserved charge densities
(which we will assume henceforward), in equilibrium%
\footnote{We use the $[{-}{+}{+}{+}]$ metric convention}
\be
T^{\mu\nu} = (\epsilon+P) u^\mu u^\nu + P g^{\mu\nu}, \qquad
u_\mu u^\mu = -1\;\;\mbox{with}\;\; u^0>0\,, \quad P=P(\epsilon) \,.
\label{T:eq}
\ee
This determines $T^{\mu\nu}$ in terms of 4 unknowns, the energy density
$\epsilon$ and 3 components of the flow 4-vector $u^\mu$.  However, if
$\epsilon,u^\mu$ vary in space and
time%
\footnote{
    It is also necessary to choose some convention defining $\epsilon$ and
    $u$.  We take the Landau-Lifshitz convention
    that $u_\mu T^{\mu\nu} \propto u^\nu$ and $\epsilon \equiv u_\mu u_\nu
    T^{\mu\nu}$.
    }
then we expect corrections to \Eq{T:eq}.  For slowly varying
$\epsilon$ and $u^\mu$ the corrections can be expanded in gradients of
these quantities.  At first order in gradients and in a conformal
theory, defining the rest-frame spatial projector
\be
\Delta^{\mu\nu} \equiv g^{\mu\nu} + u^\mu u^\nu
\ee
and working in flat space (so $\nabla_\mu=\partial_\mu$ and
$g_{\mu\nu}=\eta_{\mu\nu}$), the only possible combination is
\be
T^{\mu\nu} = T^{\mu\nu}_{\rm eq} +\Pi^{\mu\nu} \, , \quad
\Pi^{\mu\nu}_{\rm 1\;order} = - \eta \sigma^{\mu\nu} \, , \quad
\sigma^{\mu\nu} \equiv \Delta^{\mu\alpha} \Delta^{\nu\beta}
\left( \partial_\alpha u_\beta + \partial_\beta u_\alpha
- \frac{2}{3} g_{\alpha\beta} 
      \Delta^{\gamma\delta} \partial_\gamma u_\delta \right) \,.
\label{Pi_1order}
\ee
Here $\eta=\eta(\epsilon)$ is the shear viscosity, defined as the coefficient
multiplying the traceless part of the transverse symmetrized shear flow tensor.
The bulk viscosity, defined as the proportionality constant for the pure-trace
part
$\Pi^{\mu\nu} \propto \Delta^{\mu\nu} 
  \Delta_{\alpha\beta} \partial^\alpha u^\beta$,
vanishes in a conformal theory.

\Baieretal\ (\cite{Baieretal} Eq.~(3.11)) show that there are four
possible second-order flat-space terms:
\bea \hspace{-0.2in}
\label{Pi_2order}
\Pi^{\mu\nu}_{\rm 2\; order} & = &
 \eta \tau_{\Pi} \left[ u^\alpha \partial_\alpha \sigma^{\mu\nu}
                       +\frac{1}{3} \sigma^{\mu\nu} \partial_\alpha u^\alpha
                       \right]
+ \lambda_1 \left[ \sigma^{\mu}_\alpha \sigma^{\nu\alpha}-
  \frac{1}{3} \Delta^{\mu\nu} \sigma_{\alpha\beta} \sigma^{\alpha\beta}
  \right]
\nonumber \\ &&
+ \lambda_2 \left[ \frac{1}{2} (\sigma^\mu_\alpha \Omega^{\nu\alpha}
   + \sigma^\nu_\alpha \Omega^{\mu\alpha})
   -\frac{1}{3} \Delta^{\mu\nu} \sigma_{\alpha\beta}
   \Omega^{\alpha\beta} \right]
+ \lambda_3 \left[ \Omega^{\mu}{}_\alpha \Omega^{\nu\alpha}-
  \frac{1}{3} \Delta^{\mu\nu} \Omega_{\alpha\beta} \Omega^{\alpha\beta}
  \right] \,, \\
\Omega_{\mu\nu} & \equiv & \frac{1}{2} \Delta_{\mu\alpha}
    \Delta_{\nu\beta} (\partial^\alpha u^\beta - \partial^\beta
    u^\alpha) \; \mbox{ [vorticity]} \,.
 \nonumber 
\eea
Physically, $\tau_{\Pi}$ tells how quickly the anisotropic stress
 $\Pi^{\mu\nu}$ relaxes to the leading-order form $-\eta
 \sigma^{\mu\nu}$, if it starts out with a different value.
The parameter $\lambda_1$ tells how nonlinear the viscous effects are;
$\lambda_{2,3}$ are similar but for systems with nonzero vorticity.
An additional term $\kappa (R^{\mu\nu}+\ldots)$ is possible in curved
space.  It is these quantities we want to determine in weakly coupled
QCD.  We describe their physical significance in more detail
in Section \ref{sec:discussion}.

Expressing $\eta$ in terms of the dimensionless ratio $\eta/s$ disguises
the fact that $\eta$ really reports a time scale, roughly speaking the
equilibration time of the system.  The gradient expansion of
\Eq{Pi_1order}, \Eq{Pi_2order} is an expansion in this time scale
divided by the scale of spacetime variation of the system.  To identify
the time scale, divide $\eta$ not by the entropy density but by the
enthalpy density:
$\frac{\eta}{\epsilon+P} \propto \frac{1}{T}$, a time.  In 
${\cal N}{=}4$ SYM theory at strong coupling the ratio is
$\frac{\eta}{\epsilon+P} = \frac{1}{4\pi T}$.  In weakly coupled QCD it
is parametrically 
$\frac{\eta}{\epsilon+P} \sim \frac{1}{g^4 T \ln(1/g)}$
\cite{HosoyaKajantie,Baymetal,AMY1,AMY6}.  Similarly, the ratio of each
second order coefficient to $(\epsilon{+}P)$ yields the square of a time.
It is natural to expect $\frac{\lambda_1}{\epsilon+P} \sim
(\frac{\eta}{\epsilon+P})^2$.  The numerical value of the ratio 
$\frac{\lambda_1}{(\epsilon+P)} / (\frac{\eta}{\epsilon+P})^2
= \frac{(\epsilon+P)\lambda_1}{\eta^2}$ is a convenient way to express
the relative size of the second-order coefficient $\lambda_1$ to
$\eta$.  In particular we expect most coupling dependence to cancel in this
ratio, which should therefore differ relatively little between weak and
realistic coupling.

\begin{figure}
\centerbox{0.7}{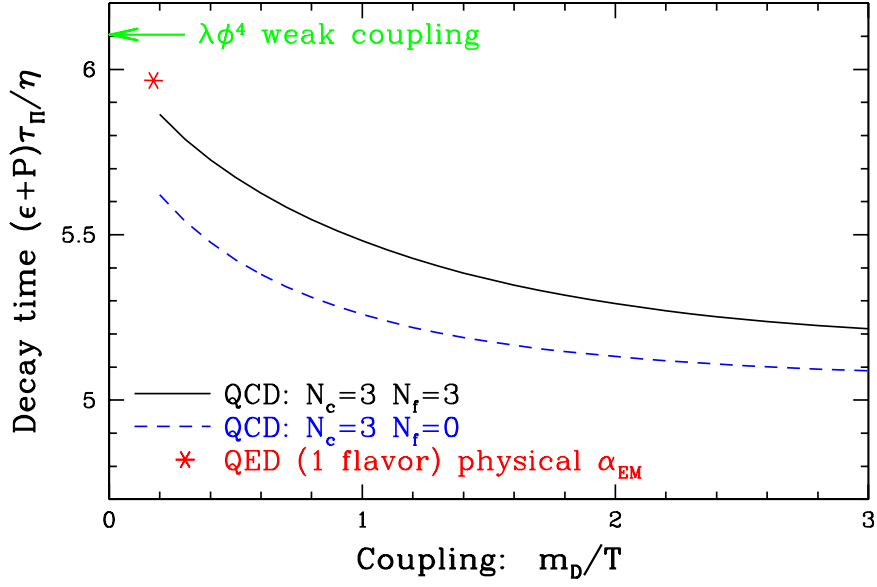}
\caption[Coupling dependence of $\tau_{\Pi}$]
{\label{fig:taupi} Coupling dependence of the ratio
$(\epsilon{+}P) \tau_{\Pi}/\eta$.  This ratio
  compares the relaxation time scale for $\Pi_{\mu\nu}$, $\tau_{\Pi}$,
  to the time scale implied by the viscosity $\eta$.
}
\end{figure}

\begin{figure}
\centerbox{0.7}{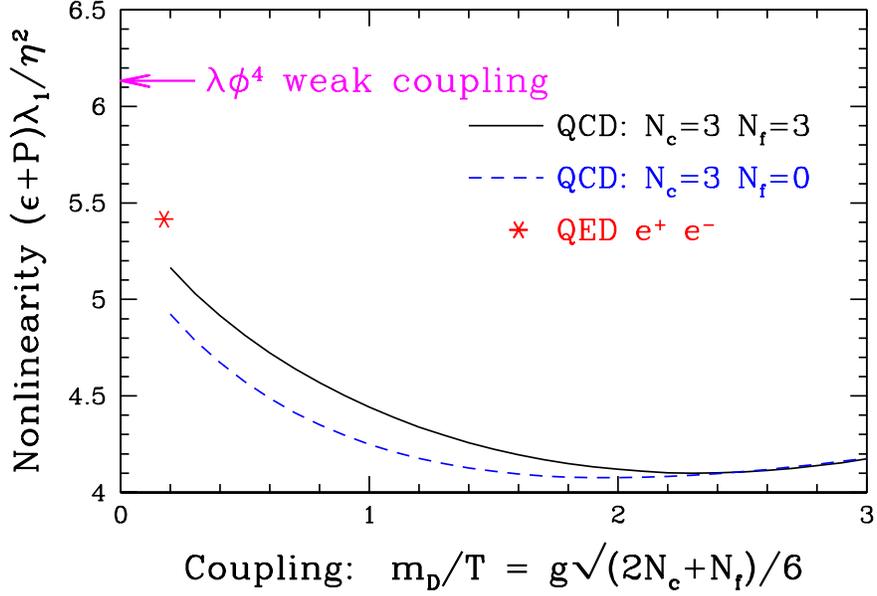}
\caption[Coupling dependence of $\lambda_1$]
{\label{fig:lambda1}
 Coupling dependence of the nonlinearity parameter $\lambda_1$,
 expressed as the dimensionless ratio
$(\epsilon{+}P)\lambda_1 / \eta^2$.
As explained in Section \ref{sec:discussion}, there is an unresolved
 uncertainty in these curves, but it is smaller than the line widths.}
\end{figure}

We find that at weak coupling, at leading order the ratios of
second-order to first-order hydrodynamic coefficients are
\bea
\label{result_tau}
\frac{(\epsilon{+}P) \eta \tau_{\Pi}}{\eta^2} & = & 
\mbox{5.9 to 5.0 (varies with $g$)}\,: \quad
6.10517 \mbox{ in } \lambda\phi^4 \mbox{ theory}\,, \\
\frac{(\epsilon{+}P) \kappa}{\eta^2} & = & 0 \, , \\
\label{result_lambda1}
\frac{(\epsilon{+}P) \lambda_1}{\eta^2} & = & 
\mbox{5.2 to 4.1 (varies with $g$)} \, : \quad
6.13264 \mbox{ in } \lambda \phi^4 \mbox{ theory}\,, \\
\frac{(\epsilon{+}P) \lambda_2}{\eta^2} & = & 
-2\frac{(\epsilon{+}P) \eta \tau_{\Pi}}{\eta^2} \, , \\
\frac{(\epsilon{+}P) \lambda_3}{\eta^2} & = & 0 \, .
\eea
The detailed coupling dependence of the two independent nonzero coefficients,
$\eta \tau_{\Pi}$ and $\lambda_1$, are shown in
Fig.~\ref{fig:taupi} and Fig.~\ref{fig:lambda1}.  These figures display
results for QCD with either 0 or 3 flavors of quarks, and for $e^+ e^-$ QED at
realistic coupling [$\frac{(\epsilon{+}P)\tau_\Pi}{\eta} = 5.9664$ and
$\frac{(\epsilon{+}P)\lambda_1}{\eta^2} = 5.4156$]
as well as indicating the results for weakly coupled
$\lambda \phi^4$ theory.  We have expressed the results in terms of $\mD/T$ the
ratio of Debye screening length and temperature, which proves convenient
computationally and is the right quantity for parametrizing whether a coupling
is strong or weak at finite temperature.  Numerically,
$\alphas = (2/12\pi) (\mD/T)^2$ in 3-flavor QCD,
$\alphas = (1/4\pi) (\mD/T)^2$ in 0-flavor QCD,
and
$\alpha_{_{\rm EM}} = (3/4\pi) (\mD/T)^2$ in 1-flavor QED.
Further discussion on these results and their physical meaning is postponed to
the discussion section, Section \ref{sec:discussion}.

We find an exact relation $\lambda_2 / \eta \tau_\Pi = -2$, in agreement
with \cite{Rischke}.  This relation is an automatic consequence of
ultra-relativistic (conformal) kinetic theory.  However unlike
\cite{Rischke} we do not find $\lambda_1 = \eta \tau_\Pi$.  This is
because \cite{Rischke} fixes an \Ansatz\ for the functional form of the
departure from equilibrium and drops some contributions arising from the
nonlinearity of the collision operator.  We discuss this in more detail
in what follows.  However in practice $\lambda_1/\eta \tau_\Pi$ is
relatively close to 1.
We also find that $\kappa = 0 = \lambda_3$ in QCD, in QED, in scalar $\phi^4$
theory, and indeed in any conformal theory described by kinetic theory.
But this does not mean that these coefficients are strictly zero; it means
that they first arise in the perturbative expansion at a higher order
than $\eta \tau_{\Pi}$ and $\lambda_1$ do.  That is, $\lambda_1 \propto
T^2/(g^8 \ln^2(1/g)) + \O (T^2/g^6)$; but $\kappa$ may only scale as,
say, $T^2/g^4$ and it is therefore zero in a leading-order evaluation, which
only finds the $\propto T^2/g^8$ coefficients.  This is discussed more
in Section \ref{sec:Kubo}.

For comparison, combining the results of \cite{Baieretal} and
\cite{Tata}, the same coefficients in ${\cal N}{=}4$ SYM theory are
\bea
\frac{(\epsilon{+}P) \eta \tau_{\Pi}}{\eta^2} & = & 
4-2\ln(2) \simeq 2.6137 \,, \\
\frac{(\epsilon{+}P) \kappa}{\eta^2} & = & 4 \, , \\
\frac{(\epsilon{+}P) \lambda_1}{\eta^2} & = & 2 \, , \\
\frac{(\epsilon{+}P) \lambda_2}{\eta^2} & = & -4\ln(2)\simeq -2.7726 \, , \\
\frac{(\epsilon{+}P) \lambda_3}{\eta^2} & = & 0 \, .
\eea
After scaling by the viscosity as described,
the second-order coefficient $\eta \tau_{\Pi}$ is about twice as large at weak
coupling as at ultra-strong coupling.  The relation between $\eta \tau_\Pi$ and
$\lambda_2$ valid at weak coupling is violated at strong coupling,
and the coefficient $\lambda_1$ is also about two times larger at weak
than at strong coupling.
It is reasonable to expect that, in QCD at realistic couplings, the
dimensionless ratios will fall between the weak-coupled values and the
(generally smaller) ultra-strong coupled SYM values.  Certainly we
expect the QCD values for these dimensionless ratios to be of the same
order of magnitude as what we find in both theories, wherever QCD is
relatively close to conformal (starting somewhat above $T_c$).
However given the difference in detail
between values in the two theories it is tough to be confident in the exact
values for realistic QCD.

\section{Kinetic theory to second order}
\label{sec:kinetic}

\subsection{Kinetic theory setup}

We will not discuss the derivation of kinetic theory here; for a review
see \cite{KadanovBaym,Lifshitz,CalzettaHu,BlaizotIancu,ASY2}.
Kinetic theory can be used when each of several criteria apply:
\begin{enumerate}
\item
There are long-lived quasiparticles (spectral functions for relevant
fields or composite operators have sharp quasiparticle peaks).
\item
The density matrix is adequately approximated by a Gaussian
approximation, that is, by the two-point function.  Further, the system
varies slowly in space and time, so we may work in terms of a space and
momentum dependent distribution function $f^a(\x,\p)$.  Here $a$ is a
label which runs over all quasiparticle types (species, spin, color,
particle/antiparticle).  (Note that $\x$
and $\p$ don't commute, but if the spatial variation is slow enough then
we can neglect the commutator and treat them as continuous, independent
variables.)
\item
The quasiparticles dominate the measurables of interest and the dynamics.
\end{enumerate}
All of these criteria hold for weakly coupled relativistic field
theories, even gauge theories \cite{BlaizotIancu}, if we are interested
in the transport coefficients which appear in the hydrodynamical
description just discussed.
The validity of the kinetic approach has been verified (at
leading order) by explicit diagrammatic analysis both in scalar field
theory \cite{Jeon} and in gauge theory \cite{Aarts,Basagoiti,Gagnon}.

The kinetic theory description describes the time
evolution of the distribution function $f^a(\x,\p)$.  This is determined
by the Boltzmann equation.  In covariant notation, it is%
\footnote{
    We use capital letters $P$ for 4-vectors, boldface $\p$ for 3-vector
    components, and $p$ for the magnitude $|\p|$ of the 3-vector.
    The collision operator here differs by a factor of $2p^0$ from
    that in \cite{Baymetal,AMY1,AMY6}.  This normalization difference will
    disappear when we integrate $\int_p$, since this integral carries a factor
    $1/2p_0$ absent in \cite{AMY1,AMY6}.  
    The overall minus sign on $\C$ is chosen
    so that its linearized form acts on the departure from equilibrium $\delta
    f$ as a positive definite operator.  To see the full covariance of the
    Boltzmann equation, think of $f^a(x,\p)$ as a function of 4-momentum $P$
    but with support only on the forward light cone, 
    $f^a(x,P) = \delta(P^2) \delta(p^0) f^a(x,\p)$.
    }
\bea
\label{eq:Boltzmann1}
2P^\mu \partial_\mu f^a(x,\p) &=& -\C[f] \mbox{ ``Collision operator''}
\\ & \equiv & -
 \sum_{a_i,b_j} \frac{1}{n_i! n_j!} \int_{k_i,k'_j}
   (2\pi)^4 \delta^4\left( P+\sum K_i - \sum K'_j \right)
   |\M|_{a_i,b_j}^2[\p,\k_i,\k'_j] \times
\nonumber \\ && \quad \qquad 
\Big( f^a(\p) \prod_{i,j} f^{a_i}(\k_i) [1{\pm}f^{b_j}(\k'_j)]
    - [1{\pm}f^a(\p)] \prod_{i,j} [1{\pm}f^{a_i}(\k_i)] 
           f^{b_j}(\k'_j) \Big)\,.
\nonumber
\eea
Here we have defined $p^0$ in terms of the on-shell condition
$p^0 = E_p\equiv\sqrt{\p^2 + m^2}=p$ (in a conformal theory $m=0$ up to
$\O(g^2)$  medium corrections, which we will neglect since we seek a
leading-order treatment),
and we have introduced the shorthand
\be
\int_k \equiv \int \frac{d^3 \k}{(2\pi)^3 2k^0}
= \int \frac{d^4 K}{(2\pi)^4} 2\pi \delta(K^2) \Theta(k^0)\,.
\ee
The lefthand side of \Eq{eq:Boltzmann1} describes the free propagation
of particles; the time rate of change of the occupancy $E \partial_t f$
is determined by the particles' motion $p^i$ times the spatial variation
of the distribution function $\partial_i f(x,p)$.  The righthand side
describes the change in occupancy due to collisions, which are
approximated as spacetime-local (so all $f$ on the RHS are evaluated at
the point $x$).  The first product of population
functions represents the rate at which particles of momentum $\p$ are
scattered out of that momentum state; $[1{\pm} f(\k')]$ is a Bose
stimulation $(+)$ or Pauli blocking $(-)$ final state factor.  The
second product of population functions is the rate for the reverse
process, producing a particle of momentum $\p$.  In equilibrium and in
the local rest frame, $[1{\pm} f_{\rm eq}(k)]=f(k) e^{k/T}$ and so the
two terms cancel by energy conservation, ensuring detailed balance.

The Boltzmann equation rests on several approximations, such as the
separation of scales between the distance between collisions 
($\O(1/g^2 T)$ in gauge theories) and the physical size of collisions
($\O(1/gT)$) or deBroglie wavelengths of excitations ($\O(1/T)$).
It is not clear how to incorporate systematic corrections
to these approximations.  It is also problematic to evaluate the
collision operator to high order in the coupling; for instance in QCD we
anticipate that nonperturbative magnetic physics causes scatterings
suppressed only by $g^2$ relative to the dominant $2\leftrightarrow 2$
scattering processes.  Indeed, we will shortly encounter (weak) logarithmic
dependence on this scale in the second-order calculation performed here.
Therefore it is not clear whether or how the
kinetic treatment can compute transport coefficients beyond leading order%
\footnote{
    Note that the first corrections to the calculations we present
    here actually arise at order $g$, not $g^2$.  However we believe
    that the $\O(g)$ corrections {\em can} be computed within
    kinetic theory; indeed this has been done in a few cases
    \cite{scalarNLO,CaronMoore}.}
in $g^2$.  So we will not try.  This excuses us to simplify the
collision operator to include only $2\leftrightarrow 2$ and
effective $1\leftrightarrow 2$ scattering processes; in QCD the relevant
collision terms are presented in \cite{AMY5}.
It also means that we can neglect the scale dependence of the QCD
coupling (the $\beta$ function).  Therefore
QCD behaves as a conformal theory,%
\footnote{For simplicity we will consider only massless QCD.}
and the analysis of \Baieretal\ \cite{Baieretal} is relevant.

\subsection{Order by order expansion}

Our goal is to solve the Boltzmann equation for the case of a
near-equilibrium system with slowly varying energy and momentum density
$(\epsilon,P^i)$, or equivalently their dual
variables, the temperature $T$ and flow velocity $u^i$.
We write $f(x,\p)$ as a formal series
\be
f(x,\p) = f_0 + \lambda f_1 + \lambda^2 f_2 + \ldots
\ee
with $\lambda$ a parameter keeping track of the order in derivatives.
The lefthand side of the Boltzmann equation, \Eq{eq:Boltzmann1} has an
explicit derivative so it starts at $\O(\lambda)$.  Therefore $f_0$ is
fixed by the condition $\C[f_0]=0$.  The solution is
(note that $u^\mu P_\mu<0$; $\beta\equiv 1/T$ as usual)
\be
f_0(x,\p) = (\exp(-\beta u^\mu P_\mu) \mp 1)^{-1} \, , \qquad
\beta=\beta(x), \; u^\mu=u^\mu(x)\, , \; p^0=p \,,
\label{eq:f0}
\ee
with $\mp=-$ for bosons and $+$ for fermions.  At first order we have
\be
2P^\mu \partial_\mu f_0 = -\C_1[f_1] \,,
\label{eq:BoltzO1}
\ee
where we use the notation $\C_1$ to mean that $\C[f]$ is expanded to
first order in $f_1$, see \Eq{eq:fexpand}.  At the second order we will have
\be
2P^\mu \partial_\mu f_1 = -\C_{11}[f_1] - \C_{1;\M_1}[f_1] - \C_{1}[f_2] \,,
\label{eq:BoltzO2}
\ee
where $\C_{11}$ is the collision operator expanded to quadratic order in
$f_1$, $C_{1}[f_2]$ is the collision operator expanded to first order in $f_2$,
and $C_{1;\M_1}$ is the collision operator expanded to first order in
$f_1$ and with the scattering matrix element also expanded to first
order in $f_1$.  In principle there could also be a term $\C_{1;m^2_1}[f_1]$
accounting for the $f_1$ dependence of particle dispersion relations, but
this will be higher order in the gauge coupling so we can ignore it in this
leading-order perturbative treatment.%
\footnote%
    {Dispersion corrections are $\O(g^2)$
    effects for the $p\sim T$ particles which dominate transport
    coefficients.  They are additionally suppressed because $f_1$ is
    chosen to have vanishing $Y_{00}(\hat\p)$ moment, and at order $g^2$ only
    this moment contributes to dispersion corrections for hard particles.
    \label{footnote:mass}
    }

It is {\em not} our goal to determine the second-order departure from
equilibrium $f_2$.  Rather, we only need to determine its contribution to the
stress-energy tensor, which at leading order in coupling is determined in
terms of $f$ by
\be
T_{\mu\nu}(x) = \sum_a \int_\p 2 p_\mu p_\nu f(x,\p)\,.
\ee
In particular this will mean that we only need spherical harmonic number 
$\ell = 2$ components of $f_2$.  However since $f_1$ appears repeatedly in the
expression \Eq{eq:BoltzO2} determining $f_2$, we need its detailed form.
Therefore the first step is to solve the first order Boltzmann equation, which
was done already in \cite{AMY6}.  So we begin by summarizing those results in
the current notation.

\subsection{First order solution}

Explicitly evaluating the lefthand side of
\Eq{eq:BoltzO1},
\be
2P^\mu \partial_\mu f_0(-\beta P\cdot u) = -
2f_0'(-\beta P\cdot u)  \Big( P\cdot u\:  P^\mu \partial_\mu \beta 
          + \beta P^\mu P^\nu \partial_\nu u_\mu \Big)\,.
\ee
Note that $f_0$ is a decreasing function so $f_0'$ is negative.
It is convenient to work noncovariantly at some point $x$ and in the
instantaneous rest frame at that point, so $u^i = 0, u^0=1$ (using Roman
letters for spatial indices, for which we will not distinguish between
covariant and contravariant).  At the point $x$ the LHS of
\Eq{eq:BoltzO1} becomes
\be
2P^\mu \partial_\mu f_0(-\beta P\cdot u) =
2f_0'(\beta E) \Big( E^2 \partial_t \beta 
    + p_i ( E \partial_i \beta - \beta E \partial_t u_i )
    - p_i p_j \beta \partial_i u_j \Big) \,.
\label{eq:1order}
\ee
Separating the spherical harmonic number $\ell=2$ and $\ell=0$
(traceless and pure-trace) parts of the last term,
\be
2p_i p_j \partial_i u_j =
\left( p_i p_j - \frac{1}{3} \delta_{ij} E^2 \right)
\left( \partial_i u_j + \partial_j u_i 
       - \frac{2}{3} \delta_{ij} \partial_k u_k \right)
+ \frac{2}{3} E^2 \partial_k u_k \,,
\ee
the $\ell=0$ contributions in \Eq{eq:1order} are
\be
2f_0' E^2 ( \partial_t \beta - \beta \partial_i u_i/3 )
\label{eq:order1ell0}
\ee
while the $\ell=1$ term is
\be
2f_0' E p_i ( \partial_i \beta - \beta \partial_t u_i )\,.
\label{eq:order1ell1}
\ee
Note that, away from equilibrium, the definitions of $\beta$ and $u^i$
are not unique; they are related to our choice of how to separate $f_0$
and $f_1$, which is also not unique.  The most sensible
convention (Landau-Lifshitz) is to require in the local rest frame
(the frame where $T^{0i}=2\sum_a \int_p p^0 p^i f^a(p)$ = 0) that the
departure $f_1+f_2+\ldots$ carry no energy or momentum,
$\sum_a \int_p p^0 P^\mu f_1^a(p)=0$.  That means choosing the
(undetermined) time derivatives $\partial_t \beta$ and $\partial_t u_i$
such that the $\int_p$ moments of the $\ell=0,1$ terms vanish.  At first
order, this requires
\be
\partial_t \beta = \frac{\beta}{3} \partial_i u_i
\qquad \mbox{and} \qquad
\partial_t u_i = \frac{1}{\beta} \partial_i \beta
\label{deltbeta}
\ee
in the instantaneous rest frame; in covariant language
\be
u^\mu \partial_\mu \beta = \frac{\beta}{3} \Delta^{\mu\nu} \partial_\mu
u_\nu
\qquad \mbox{and} \qquad
\Delta^{\nu\alpha} u^\mu \partial_\mu u_\alpha = \frac{1}{\beta} 
\Delta^{\nu\alpha} \partial_\alpha \beta \,.
\ee
This fixes the definitions of $\beta$ and $u$ at first order in
$\lambda$.  We will need these first-order relationships in evaluating
the second-order departure in what follows.
It also turns out to ensure that \Eq{eq:order1ell0} and \Eq{eq:order1ell1}
cancel identically.

This leaves the $\ell=2$ (traceless tensor) component as the sole source for
the first-order departure from equilibrium,
\be
2\beta f_0'(\beta E) 
     \left( p_i p_j - \frac{\delta_{ij} E^2}{3} \right)
     \frac{\sigma_{ij}}{2}
 = \C_1[f_1] \,,
\label{eq:BoltzO1prime}
\ee
where $\sigma_{ij}$ was introduced in \Eq{Pi_1order}.
It does not really matter whether $\sigma_{ij}$ multiplies $p_i p_j$ or
$p_i p_j - \delta_{ij} E^2/3$ in \Eq{eq:BoltzO1prime} since
$\sigma_{ij}$ projects out the trace piece; the latter shows the correct
angular behavior, the former is simpler to use in some cases.

A detailed treatment of the operator $\C_1[f_1]$ is given in
\cite{Baymetal,Jeon,AMY1,AMY6}.  What is relevant here is that
$\C_1[f_1]$ is a rotationally invariant, linear operator on $f_1$ considered
as a function of 3-momentum $\p$.  Therefore the angular
structure of $f_1$ must match that of the lefthand side;
$f_1$ must be of form
\bea
f_1(\p) &=& \frac{\sigma_{ij}}{2} (p_i p_j-\delta_{ij} E^2/3)
        \beta^3 \tchi(p) \equiv \frac{\sigma_{ij}}{2}
        \tchi_{ij}(\p)\,,
\nonumber \\
& = & \frac{\sigma_{\mu\nu}}{2} P^\mu P^\nu \beta^3 
       \tchi(-\beta u\cdot P) \qquad \mbox{(covariantly)}\,,
\label{eq:f1}
\eea
with $\tchi(p)$ a dimensionless function of $\beta$ and $p=-u_\mu P^\mu$
which remains to be determined.  By factoring out the powers of $\beta$ so
$\tchi$ is dimensionless we have ensured that it is a function only of
the dimensionless product $\beta p$ and not $\beta$ and $p$ separately.
The relation between our notation and that of \cite{AMY1,AMY6}
(AMY) is
$\tchi = \frac{T}{E^2} (-f_0') \chi_{_{\rm AMY}}$.
The departure from equilibrium $\tchi$ is generically proportional to
$-f_0' = f_0[1{\pm}f_0]$ and it is also convenient to define a version where
this is has been factored out, $\bchi = \tchi/(-f_0')$
and $\bchi_{ij} = \tchi_{ij}/(-f_0')$.
Note that $\tchi$ and $\bchi$ will both be negative definite.

It is convenient to factor out $\sigma_{ij}/2$ from both sides of 
\Eq{eq:BoltzO1prime} and to consider it as an equation on the vector space of
$\ell=2$ tensor functions of 3-momentum $\p$.  Using the inner product
\be
\langle A \, | \, B \rangle \equiv \int_p A(p) B(p)
\ee
we can define $S_{ij} = 2(p_i p_j - \delta_{ij} E^2/3)$, in which case
the first-order Boltzmann equation is
\be
\beta f_0' \,|\, S_{ij} \rangle = \C_1 \,|\, \tchi_{ij} \rangle \,.
\ee
At least formally we can then write
\be
|\, \tchi_{ij} \rangle = \beta \C_1^{-1} f_0' \,|\, S_{ij} \rangle \,.
\ee
The procedure for performing this inversion is described
in \cite{AMY1,AMY6} and here we will simply assume that this part of the
problem is already solved.  Note in particular that besides explicitly scaling
as $g^4$, the operator $\C_1$ also depends logarithmically on the coupling $g$
due to screening effects; therefore in gauge theories $\tchi(p)$ is a
nontrivial function of $g$, as is anything which functionally depends on
$\tchi$.

The first-order correction to the stress tensor is%
\footnote{
    Our $-(2E) \C_1^{-1} f_0'$ equals $\C_{_{\rm AMY}}^{-1}$
    of \cite{AMY1,AMY6}; our measure is $1/2E$ and our $S_{ij}$ is $2E$
    times the normalization used there.  These powers of $2E$ cancel
    to make the treatments equivalent.
}
\be
\Pi_{ij,{\rm 1\;order}} = \langle S_{ij} \,|\, \tchi_{lm} \rangle
\frac{\sigma_{lm}}{2} =
\frac{\sigma_{lm}}{2} \langle S_{ij} \,|\, \beta \C_1^{-1} f_0' \,|\,
S_{lm} \rangle \,.
\label{eq:eta}
\ee
In evaluating this quantity the relation for integrating over
global angles holding relative angles fixed,
\be
\frac{\sigma_{lm}}{2} \int d\Omega 
\left( \hat p_i \hat p_j - \frac{\delta_{ij}}{3} \right)
\left( \hat k_l \hat k_m - \frac{\delta_{lm}}{3} \right)
 = \frac{\sigma_{ij}}{15} P_2(\hat \p \cdot \hat \k) \,,
\label{eq:twotensor}
\ee
with $P_2(x)$ the second Legendre polynomial, is useful.

\subsection{Second order treatment}

Now we roll up our sleeves and continue to the next order.
Returning to \Eq{eq:BoltzO2}, we will find
that, formally,
\be
f_2 = -\C_1^{-1} \Big( 2P^\mu \partial_\mu f_1+\C_{11}+\C_{1;\M_1}
 \Big) \,.
\label{eq:f2}
\ee
Therefore we need to compute the three terms on the righthand side,
treating the first-order departure from equilibrium 
$\tchi(\beta E)$ as already determined.
Actually we only need to calculate that part of $f_2$ which contributes to the
off-diagonal stress tensor
\bea
\Pi_{2\;\rm order}^{ij} & = & \langle S_{ij} \,|\, f_2 \rangle
\nonumber \\
& = & - \langle S_{ij} \,|\,
       \C_1^{-1} \,|\, 2P^\mu \partial_\mu f_1
                         +\C_{11}[f_1]+\C_{1;\M_1}[f_1] \rangle \,.
\label{eq:Pi2}
\eea
But
\be
\langle S_{ij} \,|\, \C_1^{-1} 
= \langle \tchi_{ij} \,|\, (\beta f_0')^{-1} = -T \langle \bchi_{ij}|
\ee
is known; therefore we need
\be
\Pi_{2\;\rm order}^{ij} = T\: \langle \bchi_{ij} \,|\,
 2P^\mu \partial_\mu f_1
                         +\C_{11}[f_1]+\C_{1;\M_1}[f_1] \rangle \,.
\ee
In other words we need the $p$ integral, weighted with $\bchi_{ij}$, of three
terms.  No new operator inversions are required, though evaluating
$\C_{11}$ and $C_{1;\M_1}$ will require performing complicated integrals.

\subsubsection{$2P^\mu \partial_\mu f_1$ term}

We begin with the $2P^\mu \partial_\mu f_1$ term.  This contributes to
the most coefficients ($\eta\tau_\Pi$, $\lambda_1$, and $\lambda_2$) but
is the most similar to what we have already encountered.  We compute it by
evaluating $2P^\mu \partial_\mu f_1$ directly, taking the integral moment only
at the end (but feeling free to drop terms which will vanish on angular
integration).

Since we are taking its spacetime derivatives, it is necessary to use
the covariant form for $f_1$, \Eq{eq:f1}.  The derivative can act on
$\sigma^{\mu\nu}$, on $\beta$, or on $\tchi$'s
argument;
\bea
 P^\alpha \partial_\alpha \Big( \beta^3 \sigma_{\mu\nu} P^\mu P^\nu
   \tchi( -\beta u\cdot P) \Big)
& = &
\beta^3 \sigma_{\mu\nu} P^\mu P^\nu \tchi(..)
         \times 3P^\alpha \partial_\alpha \ln \beta
+ \beta^3 P^\alpha P^\mu P^\nu (\partial_\alpha \sigma_{\mu\nu}) \tchi(..)
\nonumber \\ &&
- \beta^4 \sigma_{\mu\nu} P^\mu P^\nu \tchi'(..)
   P^\alpha P^\gamma \Big( u_\gamma \partial_\alpha \ln \beta
                 + \partial_\alpha u_\gamma \Big) \,.
\eea

We only need terms which in the rest frame
are even in $\p$.
In the first and third terms $\sigma_{\mu\nu}$'s indices are spatial so
$P^\mu$ and $P^\nu$ must also be; therefore in the first term there is
only a contribution from $E \partial_t \beta$ and in the third term
there is a contribution $-E^2 \partial_t \beta$ and
$p^i p^j \partial_i u_j$ since $\partial_t u_t = 0$ at rest.
The middle term is trickier; $\partial_0 \sigma_{ij}$ can be nonzero but
so can $\partial_i \sigma_{0j}$; $\sigma_{0j}$ vanishes only at
$\vec{x}=0$ but varies from 0 away from the origin (the rest
frame at neighboring points is not the same as at the origin).  We can
re-express it using $\partial_i \sigma_{0j} = u^\beta \partial_i
\sigma_{\beta j}$, and
\be
\partial_\mu \left( \sigma^{\alpha \beta} u_\beta \right) = 
\partial_\mu ( 0 ) = 0 \quad \rightarrow \quad
u_\beta \partial_\mu \sigma^{\alpha \beta} = - \sigma^{\alpha \beta}
\partial_\mu u_\beta \,.
\ee
In other words,
\be
\partial_i \sigma_{0j} = -\sigma_{kj} \partial_i u_k \,.
\ee
Therefore the terms even in spatial indices are
(also using \Eq{deltbeta})%
\be
\label{eq:even}
\beta^3 p^i p^j E\tchi(..) \Big( \sigma_{ij} \partial_k u_k
   +  \partial_t \sigma_{ij}
   -2 \sigma_{ik} \partial_j u_k \Big)
-  \beta^3 p^i p^j \sigma_{ij} \beta \tchi'(..) \left(
    -\frac{E^2}{3} \partial_k u_k
    +p^l p^m \partial_l u_m \right) \,.
\ee
In the second term, the quantity in parenthesis is $p^l p^m
\sigma_{lm}/2$.  In the first term we need to rewrite
$\partial_j u_k$, decomposing it into its traceless symmetric,
antisymmetric, and trace components;
\bea
\partial_j u_k & = &
 \frac{\partial_j u_k + \partial_k u_j}{2}
 + \frac{\partial_j u_k - \partial_k u_j}{2}
\nonumber \\ &=&
 \frac{\partial_j u_k + \partial_k u_j - 2\delta_{jk} \partial_l u_l/3}
        {2}
 + \frac{\delta_{jk} \partial_l u_l}{3}
 + \frac{\partial_j u_k - \partial_k u_j}{2}
\nonumber \\ &=&
 \frac{\sigma_{jk}}{2} + \Omega_{jk} 
  + \frac{1}{3} \delta_{jk} \partial_l u_l \,.
\eea
Therefore this first term turns into
\be
\beta^3 \tchi(..) p^i p^j E 
\Big( \partial_t \sigma_{ij} 
  + \frac{1}{3} \sigma_{ij} \partial_k u_k
  - \sigma_{ik} \sigma_{jk}
  -2 \sigma_{ik} \Omega_{jk} \Big) \,.
\ee
This term's contribution to $\Pi_{ij,{\rm 2\;order}}$ is
\bea
\label{eq:taupi1}
\Pi_{ij,{\rm 2\; order}}
& \supset &
\Big( \partial_t \sigma_{lm} 
  + \frac{1}{3} \sigma_{lm} \partial_k u_k
  - \sigma_{lk} \sigma_{mk}
  -2 \sigma_{lk} \Omega_{mk} \Big) \times
\nonumber \\ && \qquad
\beta^5 \int_p p \left( p_i p_j - \frac{\delta_{ij} p^2}{3} \right)
\left( p_l p_m - \frac{\delta_{lm} p^2}{3} \right) \bchi(p) \tchi(p) \,.
\eea
Using \Eq{eq:twotensor} the angular integration gives $2p^5/15$, 
replacing the $lm$ indices with $ij$, removing trace parts, and leaving the
radial integral
$\beta^4 (30\pi^2)^{-1} \int pdp p^5 \bchi(p) \tchi(p)$ as the overall
coefficient.  This contributes (with negative coefficient) to $\lambda_1$
and is the sole contributor to
the terms $\tau_\Pi$ and $\lambda_2$, fixing the relation
$\lambda_2=-2\eta \tau_{\Pi}$, regardless of the form of the collision
operator (in agreement with \Baieretal\ \cite{Baieretal}). This relation
seems to be a robust prediction of kinetic theory.%
\footnote{
    This relation between $\lambda_2$ and $\tau_\Pi$ was long known
    \cite{IsraelStewart} but always in the context of Grad's 14 moment method
    \cite{Grad}; we see here that it is independent of this particular
    approximation but is more general to ultrarelativistic kinetic theory.}

Similarly, the second term in \Eq{eq:even} contributes
(note that $\bchi \tchi'<0$)
\be
\Pi_{ij,{\rm 2\;order}} \supset -2\beta^5 \int_p\tchi'
\bchi  \;
 \left( p_i p_j - \frac{p^2 \delta_{ij}}{3} \right)
 \left( p_l p_m - \frac{p^2 \delta_{lm}}{3} \right)
 \left( p_r p_s - \frac{p^2 \delta_{rs}}{3} \right) 
  \frac{\sigma_{lm} \sigma_{rs}}{4}
\,.
\ee
Evaluating this requires a special case of the angular integration relation
\Eq{eq:niftyangle}, which applied to this case gives
\be
 \int_{\Omega_{\rm global}}
 \left( p_i p_j - \frac{p^2 \delta_{ij}}{3} \right)\!\!
 \left( p_l p_m - \frac{p^2 \delta_{lm}}{3} \right)\!\!
 \left( p_r p_s - \frac{p^2 \delta_{rs}}{3} \right) 
  \frac{\sigma_{lm} \sigma_{rs}}{4}
= \frac{2p^6}{105} \left( \sigma_{il} \sigma_{jl} -
\frac{\delta_{ij} \sigma_{lm} \sigma_{lm}}{3} \right) \,.
\ee
This term contributes positively to $\lambda_1$, and is about twice as large
as the negative contribution from the first term; indeed if $\bchi$
is constant, then this factor of $2$ is exact.  Previous work
\cite{Baymetal} often used the \Ansatz\ that $\bchi$ is constant and it is not
too far from the case.  In general, if the detailed form of $\tchi$ is
known then evaluating these terms is straightforward.

\subsubsection{$C_{11}[f_1]$ term}

Now consider $\C_{11}[f_1]$.  The specific form of the collision operator now
becomes relevant; we will first consider the case of a $2\leftrightarrow 2$
collision operator.  It is convenient \cite{Baymetal} to introduce
\be
f_1(\beta E) = -f_0'(\beta E) \fbar(\beta E)
\ee
and similarly for $f_2$.
Writing $f=f_0 - f_0' (\fbar + \fbarr)$, we find to second order,
\bea
&& f(\p) f(\k) [1\ppm f(\p')] [1\ppm f(\k')] -
   [1\ppm f(\p)][1\ppm f(\k)] f(\p') f(\k') 
\nonumber \\ & = &
f_0(p) f_0(k) [1\ppm f_0(p')][1\ppm f_0(k')] \times
\nonumber \\ && \Big(
  \Big[\fbar(\p)+\fbar(\k)-\fbar(\p')-\fbar(\k')\Big]
+ \Big[\fbarr(\p)+\fbarr(\k)-\fbarr(\p')-\fbarr(\k')\Big]
\nonumber \\ && \phantom{\Big(}
+ \fbar(\p)\fbar(\k) f_0(p) f_0(k) (e^{\frac{p+k}{T}}-1)
+ \fbar(\p')\fbar(\k') f_0(p') f_0(k') (1-e^{\frac{p+k}{T}})
\nonumber \\ && \phantom{\Big(}
+\left[ \fbar(\p) \fbar(\p') f_0(p) f_0(p') 
                               \left(e^{\frac{p}{T}}-e^{\frac{p'}{T}}\right)
+ (p'\rightarrow k')
+ (p \rightarrow k)
+ (p,p' \rightarrow k,k') \right] \Big) \qquad
\label{eq:fexpand}
\eea
plus terms which are third order in gradients.
Here $(p'\rightarrow k')$ means the first term in the square brackets, but
with the substitution $p'\rightarrow k'$.
The first two square-bracketed terms are responsible for $\C_1[f_1]$
and $\C_1[f_2]$; the last two lines
are quadratic in $f_1$ and are therefore what we meant by $\C_{11}$ terms.
The contribution of $\C_{11}$ to $\Pi_{ij}$ will involve
\bea
\label{eq:C11}
\Pi_{ij,{\rm 2\;order}} & \supset &
\int_{pkp'k'} (2\pi)^4 \delta^4(P{+}K{-}P'{-}K')
|\M|^2 f_0(p) f_0(k) [1{\pm} f_0(p')][1{\pm}f_0(k')] \times
\\ &&
T \bchi_{ij}(\p) \frac{\sigma_{lm} \sigma_{rs}}{4}
\left[
\bchi_{lm}(\p) \bchi_{rs}(\k) f_0(p) f_0(k)
\left( e^{\frac{p+k}{T}} - 1 \right) + \vphantom{\Big|}
\mbox{ 5 more terms} \right] \,.
\nonumber
\eea

For collinear effective $1\leftrightarrow 2$ processes we similarly need
($p'+k'=p$)
\bea
&& f(\p) [1\ppm f(\p')] [1\ppm f(\k')] -
   [1\ppm f(\p)] f(\p') f(\k') 
\nonumber \\ & \!=\! &
f_0(p) [1\ppm f_0(p')][1\ppm f_0(k')] \times
\\ && \Big(
  \Big[\fbar(\p)-\fbar(\p')-\fbar(\k')\Big]
+ \Big[\fbarr(\p)-\fbarr(\p')-\fbarr(\k')\Big]
\nonumber \\ && 
+ \fbar(\p')\fbar(\k') f_0(p') f_0(k') (1-e^{\frac{p}{T}})
+\left[ \fbar(\p) \fbar(\p') f_0(p) f_0(p') 
                       \left(e^{\frac{p}{T}}-e^{\frac{p'}{T}}\right)
+ (p'\rightarrow k') \right]
 \Big)\,. \nonumber
\eea
The contribution to $\Pi_{ij}$ is of similar form to \Eq{eq:C11}.  These terms
clearly depend in detail on the available processes and their matrix elements
$|\M|^2$; they also require multi-dimensional integration over the external
particle momenta.  However the relevant matrix elements and useful
parameterizations for the angular integrations have already appeared
\cite{AMY6}, so we will concentrate on what is new, which is the angular
structure.

In evaluating \Eq{eq:C11} we will encounter an integration over
global angles, keeping relative angles between $\p,\p',\k,\k'$ fixed.
Since the matrix elements do not depend on global angles, we may perform this
global angular integration first.  Introducing the notation
\be
\hat\p_{\langle i} \hat\q_{j\rangle} \equiv
\frac{1}{2} \left( \hat\p_i \hat\q_j + \hat\q_i \hat\p_j
-\frac{2}{3} \delta_{ij} \hat\p\cdot \hat\q \right)
\ee
for the traceless symmetrized part, the generic integral we need
is of form
\be
\frac{\sigma_{lm} \sigma_{rs}}{4} \int_{\Omega_{\rm global}} 
\hat\p_{\langle i} \hat\p_{j\rangle}
\hat\k_{\langle l} \hat\k_{m\rangle}
\hat\p_{\langle r}' \hat\p_{s\rangle}' \,,
\ee
where we will normalize so that $\int_{d\Omega_{\rm global}} 1 = 1$.
We show how to deal with a slight generalization of this form, needed in
evaluating $C_{1;\M_1}$.  Consider
\be
\frac{\sigma_{lm} \sigma_{rs}}{4} \int_{\Omega_{\rm global}}
A_{ij} B_{lm} C_{rs} \, , \qquad
A,B,C \mbox{ of form } A_{ij} = \hat\p_{\langle i} \hat\q_{j\rangle}
\ee
that is, each $A,B,C$ is a distinct traceless symmetric tensor.  The
global angular integration over $A_{ij} B_{lm} C_{rs}$ must give a
rank-6 tensor, symmetric and traceless on each pair of indices.  There
is only one such tensor:
\bea
\int_{\Omega_{\rm global}} A_{ij} B_{lm} C_{rs}
&=& C[A,B,C] \left( \delta_{il} \delta_{jr} \delta_{ms} +
\mbox{7 permut.}
-\frac{4}{3} ( \delta_{rs} \delta_{il} \delta_{jm}
+\mbox{5 permut.} )
\right. \nonumber \\ && \qquad \qquad \qquad \left.
+ \frac{16}{9} \delta_{ij} \delta_{lm} \delta_{rs} \right) \,.
\eea
The coefficient $C[A,B,C]$ is determined by contracting each side with
$\delta_{il} \delta_{jr} \delta_{ms}$, yielding
\be
C[A,B,C] = \frac{3}{70} A_{ij} B_{im} C_{jm} \,.
\ee
Therefore 
\be
\frac{\sigma_{lm} \sigma_{rs}}{4} \int_{\Omega_{\rm global}}
A_{ij} B_{lm} C_{rs}
= \frac{3}{35}\left( \sigma_{il} \sigma_{jl} - \frac{\delta_{ij}}{3}
 \sigma_{lm} \sigma_{lm} \right) A_{rs} B_{rt} C_{st} \,.
\label{eq:niftyangle}
\ee
In particular, in evaluating \Eq{eq:C11} we will need angular
moments of form
\bea
&&
\frac{\sigma_{lm} \sigma_{rs}}{4} \int_{\Omega_{\rm global}} 
\hat\p_{\langle i} \hat\p_{j\rangle}
\hat\k_{\langle l} \hat\k_{m\rangle}
\hat\p_{\langle r}' \hat\p_{s\rangle}'
\nonumber \\
& = & \frac{1}{35}
\left( \sigma_{il} \sigma_{jl} - \frac{\delta_{ij}}{3}
 \sigma_{lm} \sigma_{lm} \right)
\left( 3 x_{pk} x_{pp'} x_{kp'} - x_{pk}^2 - x_{pp'}^2 - x_{kp'}^2
+\frac{2}{3} \right)
\label{nifty_angle}
\eea
where we define $x_{pk} = \hat\p\cdot \hat\k$.  This result together
with results in \cite{AMY6} are sufficient to compute the $\C_{11}$
contribution.  Note that the contraction of $\sigma$ tensors above is
precisely the one defining the coefficient $\lambda_1$ in
\Eq{Pi_2order}.  Therefore the term $\C_{11}$ strictly contributes to
$\lambda_1$.

\subsubsection{$C_{1;\M_1}$ contribution}

If we calculated $\tchi$ in a gauge theory, using the
vacuum matrix elements, we would find a log divergence in $\C_1$ due to the
Coulomb singularity, and therefore $\tchi$ would be zero.  Therefore it is
essential in applying kinetic theory in a gauge setting to include the physics
of dynamical screening \cite{Baymetal}, both for gauge boson and for fermion
exchange.

However, dynamical screening depends on the density of plasma particles and
their momentum distribution; the matrix element $\M$ is itself
a function of $f$, $\M[f]$.  Since $f=f_0 + f_1 + \ldots$, we can expand the
matrix element as well;
\be
\M[f] = \M[f_0] + \lambda \int_r f_1(r) \frac{d\M[f]}{df(r)} + \O(\lambda^2)
\ee
where as before $\lambda$ keeps track of orders in gradients.
As shown in \Eq{eq:fexpand}, the product of population functions in the
collision operator is only nonzero at $\O(\lambda)$; therefore the
$\O(\lambda)$ correction to $\M$ first gives rise to a nonzero effect at
second order in $\lambda$.  In particular
\bea
\label{eq:C1M1}
\C_{1;\M_1}[f_1] & = & \int_{kp'k'} (2\pi)^4 \delta^4(\ldots)
\int_r f_1(r) \left( \M[f_0] \frac{d\M^*[f]}{df(r)} + {\rm h.c.} \right)
 \\ && \qquad \times
f_0(p) f_0(k) [1{\pm}f_0(p')][1{\pm}f_0(k')]
\Big( \fbar(\p) + \fbar(\k) - \fbar(\p') - \fbar(\k') \Big)\,.
\nonumber
\eea
The contribution to $\Pi_{ij}$ is $\int_p T \bchi_{ij}$ of this.

The functional form of $d\M/df$ is somewhat complicated but is only
significant for small exchange momenta, that is, when one of the Mandelstam
variables is small, say, $t \lsim \mD^2$.  Therefore, in the context of a
perturbative treatment it is fair to work in the small exchange momentum
approximation, $|t| \ll s$.  This simplifies both the form of $d\M/df$ and of
the integration structure.  However the specific details for evaluating
$\C_{1;\M_1}$ are complicated enough that we have postponed them
to Appendix \ref{sec:appendix}.


\section{Kubo formula for $\tau_{\Pi}$ and $\kappa$}
\label{sec:Kubo}

The previous discussion has determined all but one of the second-order
hydrodynamic coefficients; since we worked in flat space we were unable to
determine the coefficient $\kappa$.  Here we evaluate $\kappa$ without leaving
flat space, and provide an alternative evaluation of $\tau_\Pi$, by making use
of a Kubo relation derived
by \Baieretal\ \cite{Baieretal}.  There it is shown that the two ``linear''
second-order coefficients, $\eta \tau_{\Pi}$ and $\kappa$, can be
determined if one can evaluate the retarded Green function for the
stress tensor%
\footnote
    {Our convention for the retarded function is missing a factor of $i$
      found in many definitions; our retarded function for a free
      particle is $G_{\rm R}(P)=-i/(P^2+m^2+i\epsilon p^0)$
      or $G_{\rm R}(P) =i/(p^0-E+i\epsilon)(p^0+E+i\epsilon)$.
      Therefore $2{\rm Disc}\:G_R(\omega)=\rho(\omega)$
      the spectral function is real.}
\be
G_{\rm R}^{T_{xy} T_{xy}}(\omega,\k) \equiv
\int d^4 x e^{-i\omega t + i\k\cdot \x} \Theta(t)
\Tr \rho_T \commut{T_{xy}(0)}{T_{xy}(x)}
\ee
(with $\rho_T$ the equilibrium, thermal density matrix)
and expand it to second order in $\omega,k_z$ at vanishing $k_x,k_y$.
In particular (Eq. (3.14) of \cite{Baieretal} in our conventions)
\be
G_{\rm R}^{T_{xy} T_{xy}}(\omega,k_z) = -i P + \eta \omega
+i \left( \omega^2 (\eta \tau_{\Pi}-\kappa/2) - k_z^2 \kappa/2 \right)
      \,.
\ee
Note that all correlation functions in this section are for a plasma in
equilibrium.

We can use kinetic theory to compute a related equilibrium
correlator, the Wightman function
\be
G^{>,T_{xy} T_{xy}}(\omega,k) \equiv
\int d^4 x e^{-i\omega t + i\k\cdot \x}
\Tr \rho_T T_{xy}(0)T_{xy}(x) \,.
\ee
The relation between these correlation functions is that
\be
G^>(\omega,k) = \frac{1}{1-e^{-\omega/T}} 
      ( G_R(\omega+i\epsilon) - G_R(\omega-i\epsilon) )
\simeq \frac{T}{\omega} 2 \Re G_R(\omega+i\epsilon) \,.
\ee
(In the second relation we made the approximation $\omega \ll T$, valid
for all frequencies of relevance here.)  This relation can be inverted
into a Kramers-Kronig relation
\be
G_{\rm R}(\omega') = -i \int \frac{d\omega}{2\pi} 
       \frac{1}{(\omega - \omega' -i\epsilon)} \;
       \frac{\omega}{T} G^>(\omega) \,.
\ee

To evaluate the Wightman function $G^>$, recall that the Fermi/Bose
distributions have fluctuations which are independent for each $a,\p$
and of magnitude $\delta f(p) = f_0[1{\pm} f_0]=-f_0'(p)$.
The instantaneous value of $T_{xy}$ is
\be
T_{xy}(x,t) = 2\int_p p_x p_y \delta f(p,x,t) \,,
\ee
which averages to zero.  But the two-point function does not;
\be
G^>(x,t)= \langle T_{xy}(0,0) T_{xy}(x,t) \rangle 
 = 4\int_{pp'} p_x p_y p'_x p'_y 
   \langle \delta f(p',0,0) \delta f(p,x,t) \rangle \,.
\ee
We can evaluate this at positive $t$ by pretending that $p'_x p'_y
f_0[1{\pm}f_0](p')$ is a source for departure from equilibrium in the
Boltzmann equation and evaluating the expectation value for $T_{xy}$
with the resulting departure linearized%
\footnote{
    $\delta f$ is not quite the same as $f_1$ in the previous section; it
    includes terms second order in gradients but first order in the departure
    from equilibrium, that is, it will contain terms quadratic and higher in
    spacetime derivatives but is linear in $u_i$.
    }
$\delta f(p,x,t)$.  The relevant Boltzmann equation is
\be
2p_x p_y f_0'(p) \delta(t) \delta^3(x)
+ 2(E \partial_t + p_i \partial_i) \delta f_1(p,x,t) 
    = -\C_1[\delta f(x,t)] \,.
\ee
The spatial Fourier transform is trivial, removing $\delta^3(x)$ and
replacing $\partial_i$ with $ik_i$.  The time transform is more subtle.
If $\C$ were replaced by a relaxation time 
$\C[f] \rightarrow 2E \Gamma f_1$ and ignoring $k_z$ for the moment, we would
have
\bea
\delta f[\mbox{relax-time-approx};t] &=& \frac{e^{-\Gamma |t|}}{2E} 
 (-f_0')\: 2 p_x p_y  \,,
\nonumber \\
\delta f[\mbox{relax-time-approx};\omega] & = & \left( 
     \frac{1}{2E\Gamma + 2i\omega E} + \mbox{c.c.} \right)
     (-f_0')\: 2p_x p_y \,.
\eea

Instead $\C$ is an operator.  Moving the spacetime derivatives to the
righthand side and formally inverting, one finds
\be
|\delta f(\p;\omega,k_z)\rangle = \left(
  \frac{1}{\C+2i(\omega E-k_z p_z)} + \mbox{c.c.} \right) (-f_0')
|S_{xy}\rangle \, .
\ee
The stress-stress correlator is the value of $T_{xy}$ arising from this $f_1$,
which is
\be
G^>(\omega,k_z) = \langle S_{xy} | \delta f \rangle
 = \langle S_{xy} | \left( \frac{1}{\C+2i(\omega E - k_z p_z)}
        +\mbox{c.c.} \right) f_0' |S_{xy} \rangle \,.
\ee

Now we use $G^>$ and the Kramers-Kronig relation to evaluate $G_{\rm R}$.
First consider the case where $k_z=0$ but we allow $\omega'$ to be
finite.  Then (combining fractions)
\be
G_{\rm R}(\omega') = \langle S_{xy} | \int \frac{-i d\omega}{2\pi T}\:
   \frac{\omega}{(\omega-\omega'-i\epsilon)} \:
 \frac{2\C}{(\C+i2E\omega)(\C-i2E\omega)} (-f_0') | S_{xy} \rangle \,.
\ee
Because $\C$ has a purely real and
positive spectrum, we are free to perform the $\omega$ integral by the
method of residues, enclosing only the pole arising from
$(\C-i2E\omega)$;
\be
G_{\rm R}(\omega') = -i\langle S_{xy}| \frac{1}{2ET} \:
       \frac{\C}{\C-i2E\omega'} (-f_0')
| S_{xy} \rangle 
=\sum_{n=0}^{\infty} -i \langle S_{xy}| \frac{1}{2ET} 
   \left( 2iE\omega \C^{-1} \right)^n (-f_0') |S_{xy}\rangle \,.
\label{eq:Gret}
\ee
The leading term in the expansion is
\be
-i\langle S_{xy}| \frac{1}{2ET} (-f_0') |S_{xy}\rangle
= -i \sum_a \int \frac{d^3 \p}{(2\pi)^3 T} \frac{p_x^2 p_y^2}{p^2}
   f_0[1{\pm}f_0]
=-i g_* \frac{4}{5} \frac{\pi^4 T^4}{90}
\ee
which is $\frac 45$ of the expected $-iP$.  (Here 
$g_* = \sum_a (1\mbox{(boson) or }\frac{7}{8} \mbox{ (fermion)})$.)
The remaining $\frac 15$ of $-iP$ arises from
$\omega\simeq T$ (large frequency cut) contributions to $G^>$ which we have
not computed here, and which give only order $g^0$ contributions to
$\eta,\eta\tau_\Pi,\kappa$, which we therefore neglect.

The first subleading $\propto \omega'$ term in \Eq{eq:Gret}
reproduces \Eq{eq:eta}
and the last term allows us to calculate the
combination $(\eta \tau_\Pi - \kappa/2)$:
\be
\eta \tau_\Pi - \kappa = 
\beta \langle S_{xy}\,|\, \C_1^{-1} (2E) \C_1^{-1} (-f_0')
\,|\, S_{xy} \rangle
= T \langle \tchi_{xy} \,|\, 2E (-f_0') \,|\, \tchi_{xy} \rangle
\ee
which leads to the same result we had for $\eta \tau_\Pi$ previously
in \Eq{eq:taupi1}.  This already shows us that $\kappa=0$.

To establish that $\kappa=0$ in another way, we
directly evaluate the second order in $k$ term at vanishing $\omega'$.
The retarded Green function is
\be
G_{\rm R}(\omega'=0,k_z) = \langle S_{xy} |
  \int \frac{-i d\omega}{2\pi T}  \frac{\omega}{\omega-\omega'-i\epsilon}
\left( \frac{1}{\C_1-i2E \omega +i2p_z k_z} +\mbox{c.c.} \right) 
         (-f_0') |S_{xy}\rangle \,.
\ee
The ratio $\omega/(\omega-\omega'-i\epsilon)$ cancels.%
\footnote{
    The integrand needs to be regular at $\omega=0$ for this cancellation to
    work, otherwise the $i\epsilon$ prescription is nontrivial.  However the
    good properties of $\C_1$ ensure this is the case.
    }
Because $\C$ has positive definite spectrum we can again perform the
$\omega$ integral by closing the contour above for the
$1/(\C-iE\omega+ip_z k_z)$ term and below for the
$1/(\C+iE\omega-ip_z k_z)$ term.  There are no poles to pick up, but
there is a nonzero contribution from the contour-closing arc because the
integrand only falls as $1/\omega$.  However this arises in the extreme
large $\omega$ region where the finite operators $\C,p_z$ are subdominant and
can be dropped.  Therefore we find a $k_z$ independent result.  Equivalently,
we could Taylor expand about small $k_z$,
\be
\frac{1}{\C+i2E\omega-2ip_z k_z}
= \frac{1}{\C+i2E\omega}
+ \sum_{n=1}^{\infty} \frac{1}{\C+2iE\omega}
\left( 2ip_z k_z \frac{1}{\C+2iE\omega} \right)^n
\ee
and integrate term by term; on all but the first $k_z$ independent term
the integrand falls as $1/\omega^2$ or faster, and we may close the
contour away from all poles and pick up no contribution.

Therefore the expansion of $G_{\rm R}(\omega,k)$ in powers of $k_z$ at
vanishing $\omega$ shows no $k$ dependence, and the second-order
coefficient $\kappa$ vanishes.  To clarify, the expansion in nonzero
$k_z$ {\em and} $\omega$ will contain nonvanishing terms, of order
$\omega k_z^2$ {\it etc}.  It is only the $k_z$ dependent terms at
$\omega=0$ (or vanishing order in $\omega$) which vanish
in kinetic theory.  Note that we did not have to make
{\em any} assumptions about the collision operator $\C$ to arrive at
this conclusion, except that it is space-local and positive definite
(the equilibrium ensemble is stable against perturbations).

This result is not too surprising.  As explained in \cite{Baieretal},
another way of interpreting the $k_z^2$ coefficient is that it gives the
correction to the stress tensor if there is a spatially varying but
time-independent traceless metric disturbance $h_{xy}(z) \neq 0$.  But
examining classical phase-space trajectories for this specific background
shows that an initially equilibrium distribution freely propagates to remain
in equilibrium (at linearized order and when the geometry is time
independent).  Explicitly, in curved space the Boltzmann equation
is \cite{Nabeel}
\be
p^\mu \partial_{x^\mu} f(x,p) - \Gamma^\lambda{}_{\mu\nu}
p^\mu p^\nu \partial_{p^\lambda} f(x,p,t) = -\C_1[\delta f] \,.
\label{eq:Boltzcurved}
\ee
For the case
$g_{\mu\nu} = \eta_{\mu\nu} + h_{\mu\nu}$,
$h_{xy} = h_{yx} = \alpha e^{ikz}$ with $\alpha$ time independent and all
other components zero, the nonzero Christoffel symbols are
\be
\Gamma^x{}_{yz} = \Gamma^y{}_{xz} = -\Gamma^z{}_{xy} = 
\frac{1}{2} \partial_{x^z} h_{xy} \,.
\ee
Since $\Gamma$ is already linear in $h$ we may evaluate 
$\partial_{p^\lambda} f$ using the flat-space form for $f_0$,
$\partial_{p^\lambda} f_0 = (f_0') p^\lambda/p$.  The second term on the
lefthand side of \Eq{eq:Boltzcurved} is therefore
\be
-\Gamma^\lambda{}_{\mu\nu} p^\mu p^\nu \partial_{p^\lambda} f_0
= - \frac{p^x p^y p^z}{p} (f_0') \partial_{x^z} h_{xy} \,.
\ee

To evaluate the first term, we have to evaluate $f$ to first order in $h$.
The equilibrium form is 
$f_0 = (\exp(\beta g_{\mu\nu} u^\mu P^\nu) \mp 1)^{-1}$,
and since only $u^0$ is nonzero and $g_{0\nu}$ is unchanged this is
$f_0 = 1/(e^{-\beta p^0} \mp 1)$.  However $p^0$ is defined implicitly in
terms of $p^i$ via
$g_{\mu\nu} P^\mu P^\nu = 0$.  
Therefore
$p^0 = \sqrt{\p^2 + 2h_{xy} p^x p^y} = p + h_{xy} p^x p^y/p$ plus terms
quadratic in $h$.  Evaluating the space derivative therefore gives
\be
p^\mu \partial_{x^\mu} f_0 = \frac{p^z p^x p^y}{p} (f_0')
\partial_{x^z} h_{xy} \,.
\ee
The two terms cancel, meaning that the system remains exactly in equilibrium
to linearized order in $h$.

Since this argument relies only on classical phase space propagation, the
coefficient $\kappa$ will first arise when this classical phase-space
picture becomes insufficient.  The parametric behavior of $\lambda_1 \sim
T^2/g^8$ arose as $T^4/l_{\rm mfp}^2$, involving two powers of the mean free
path.  Our phase space argument shows that $\kappa$ must involve one power of
the scale where classical phase space treatments break down, which is the
scale set by the inverse deBroglie wavelength $T$.
Therefore we expect that $\kappa \sim T^4/(l_{\rm mfp} T) \sim T^2/g^4$ (at
most).  Computing the first nonvanishing contributions to $\kappa$ at weak
coupling is beyond the scope of kinetic theory and of this work.

\section{Discussion}
\label{sec:discussion}

We clarify and discuss in turn the meaning and origin of the
second-order coefficients within kinetic theory.  In particular, consider
shear flow with $\sigma_{zz}=-2c$, $\sigma_{xx}=\sigma_{yy}=c$ with $c$
positive.  This is Bjorken contraction, with some radial expansion to preserve
volume (or pure Bjorken contraction plus a conformal transformation).
In this case we expect a particle distribution to become prolate along the $z$
axis, leading to $T_{zz}>T_{xx},T_{yy}$.  This is what happens.  The
magnitude, integrated over $p^3 dp$, determines $\eta$.  The deviation
from equilibrium depends on $p$ and is described by $p^2\bchi(p)$, the
relative departure from equilibrium $f_1/f_0[1{\pm}f_0]$ as a function of $p$.
In a relaxation time approximation, $\bchi \propto 1/p$; in a momentum
diffusion approximation $\bchi \propto 1$.

The physical meaning of $\tau_\Pi$ is, how far $T_{zz}$ comes from this
expected form if the rate of Bjorken contraction is changing with time.
If Bjorken contraction is speeding up, the particle distribution should reflect
the smaller value which used to be valid; hence $T_{zz}$ should be smaller,
meaning the proportionality constant $T_{zz}=-\eta \sigma_{zz}
+\eta\tau_\Pi \partial_t \sigma_{zz}$ should be positive (since $\sigma_{zz}$
is negative).  This is the sign we obtain.  But how much smaller?  This
depends on how quickly the distribution relaxes back to equilibrium.  The size
of $\eta/(\epsilon{+}P)$ also depends on how quickly the distribution relaxes
to equilibrium, so we expect some relationship
$\tau_\Pi \sim \eta/(\epsilon{+}P)$.  But the proportionality constant
depends on whether all particles
equilibrate in the same way, or some particles take longer to equilibrate.  If
high momentum particles take longer to relax to equilibrium, then they can
store information about the value of $\sigma_{zz}$ further into the past.  As
a result, if we make a relaxation time approximation, then
$\bchi\propto 1/p$ gives $\tau_\Pi = 5\eta/ (\epsilon{+}P)$, whereas
the momentum diffusion approximation
$\bchi \propto 1$ gives $\tau_\Pi = 6 \eta/ (\epsilon{+}P)$.  Figure
\ref{fig:taupi} shows that the value moves from close to 6, at weak coupling,
to nearly 5 at stronger coupling.  This occurs because at weak coupling
collisions
are dominated by soft scattering, which acts like momentum diffusion and gives
quite close to $\bchi\propto 1$ (see \cite{AMY1}), while at larger couplings
collinear splittings become more important and try to enforce
$\bchi \propto 1/p$ (see \cite{AMY6}).  So this coupling behavior is
expected.%
\footnote{
    The value in scalar $\lambda\phi^4$ theory is slightly higher than 6.
    However, if we replace Bose statistics with Boltzmann statistics, it turns
    out that the \Ansatz\ $\bchi \propto 1$ is exact, and
    $\tau_{\Pi}(\epsilon{+}P)/\eta = 6$ exactly at leading order in $\lambda$.
    }

\begin{figure}[b]
\centerbox{0.6}{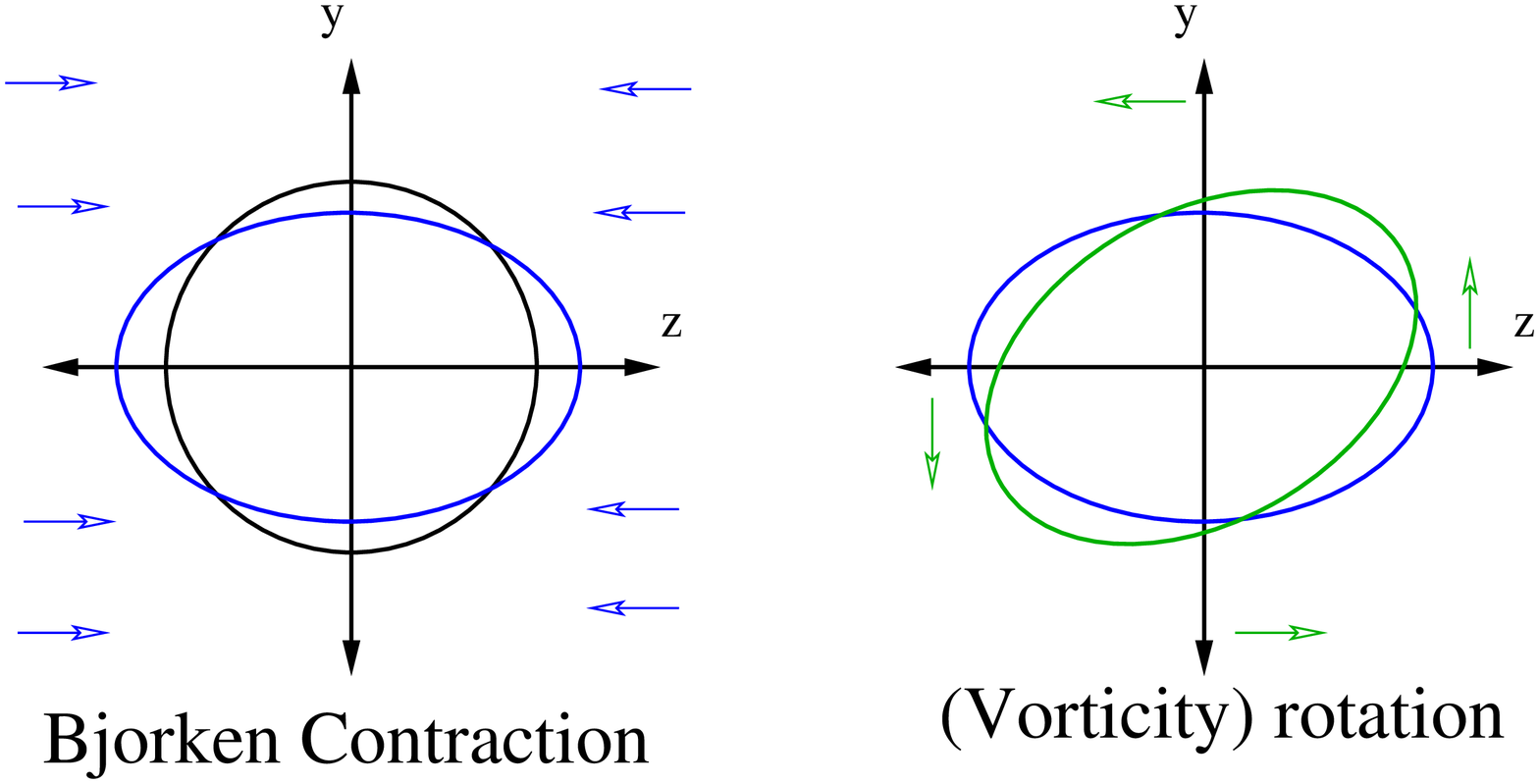}
\caption[Explaining $\lambda_2$]
{\label{fig:lambda2}
 Illustration of the physical origin of $\eta$ and of $\lambda_2$.  Under
 Bjorken contraction (left), the momentum distribution becomes prolate long
 the $z$ axis.  But under rotation with $\partial_z v_y > 0$ (right), the
 prolate axis gets rotated to have a $y$ component, so $T_{yz}>0$.
}
\end{figure}

The relation between $\tau_\Pi$ and $\lambda_2$, and the sign of $\lambda_2$,
also have fairly simple interpretations.  First the sign.  Physically
$\lambda_2$ tells what happens to a system which is both Bjorken contracting
(nonzero $\sigma_{zz}$) and rotating (nonzero vorticity, say
$\Omega_{zy}>0$).  As illustrated in Figure \ref{fig:lambda2}, in this case
the contraction makes the particle distribution become prolate; but the
vorticity skews this distribution so it is not aligned with the Bjorken
contraction axis.  That should lead to a positive $T_{yz}$,
which for $\sigma_{zz}<0$ and $\Omega_{zy}>0$ requires $\lambda_2<0$.
The proportionality constant depends on how large the original $zz$ asymmetry
was, which depends on $\eta$, and on how long the induced $xy$ skewed
distribution ``lives,'' which is set by $\tau_\Pi$.  Accounting for numerical
factors turns out to give $\lambda_2 = -2\eta \tau_\Pi$, as we find.  

Next consider $\lambda_1$.  For our example of Bjorken contraction,
$$
\Pi_{zz} = -\eta \sigma_{zz} + \lambda_1 (\sigma_{zl} \sigma_{zl}
-\delta_{zz} \sigma_{lm}^2/3) = \eta (2c) + \lambda_1 (2c^2) \,.
$$
Therefore a positive $\lambda_1$ means that for Bjorken contraction,
the stress tensor deviates further than normal from equilibrium.
On the other hand, reversing the sign of $c$ to consider Bjorken expansion,
the deviation from the equilibrium value of $\Pi_{zz}$ is reduced.
Therefore $\lambda_1$ tells whether equilibration is accelerated
for Bjorken expansion ($\lambda_1$ positive) or Bjorken contraction
($\lambda_1$ negative).%
\footnote%
    {
     $\lambda_1$ does NOT indicate the ``anomalous
     viscosity'' expected from plasma instabilities \cite{BassMuller}.
     ``Anomalous viscosity,'' for which
     $|\Pi_{ij}|$ falls below the linear term for all flow patterns,
     would be indicated by a large positive value for the third order term
     $\Pi_{ij} \propto \sigma_{ij} \sigma_{lm} \sigma_{lm}$.
    }

Our calculation shows that there are 3 contributions to $\lambda_1$.  First,
if the particle distribution has already become prolate, then further Bjorken
contraction generates a different amount of prolateness than it would from a
spherically symmetric distribution.  This is the part contributed
by $2P^\mu \partial_\mu f_1$.  The sign turns out to be positive and the
magnitude dominates all contributions to $\lambda_1$.

The contribution to $\lambda_1$ from $\C_{11}$ reflects the change, in going
from a thermal to a prolate momentum distribution, in the set of scattering
targets a particle has.  Whether this accelerates equilibration or slows it
down depends on typical scattering angles in a rather complicated way,
indicated by the rather complicated angular integrations involved
in \Eq{nifty_angle} and \Eq{eq:C11}.  This leads to considerable angular
cancellation.  For instance, in $\lambda \phi^4$ theory, where the matrix
element $\M^2 = \lambda^2$ shows no preference for particular scattering
angles, the contribution to $\lambda_1$ from $\C_{11}$ is $+0.0372$.  If we
replace Bose with Boltzmann statistics in $\lambda \phi^4$ theory,
the cancellation on angular averages becomes exact and $\C_{11}$ gives 
{\it no} contribution to $\lambda_1$.  In QCD the contribution is also small,
due to significant angular cancellation; for 3-flavor QCD
the $\C_{11}$ contribution to $\lambda_1$
varies between $-0.18$ at weak to $-0.45$ at stronger coupling.  The
negative sign means that prolate distributions show accelerated equilibration.

The contribution to $\lambda_1$ from $\C_{1;\M_1}$ reflects changes in the
efficiency of scattering and collinear splitting because of changes in plasma
screening.  This is interesting because it is where the precursors of plasma
instabilities (see \cite{Mrow,RRS,ALM}) can enter the game.  An anisotropic
particle distribution weakens the stabilizing effect of plasma screening for
certain particle directions $\hat{p}$ and exchange momenta $\q$.
In particular, in directions where $f_1(\p)$ is positive,
these particles have enhanced scattering
via soft magnetic ($G_T$) gluon exchange with $\q \perp \p$.
One might guess that this leads to a large negative contribution to
$\lambda_1$.  However we find that extensive angular cancellations occur which
make the contribution arising from elastic scatterings very small, and free of
IR divergences, see the discussion at the end of Appendix
\ref{sec:C1m1bos}.

The same does not happen for collinear splitting.  If the particle
distribution becomes prolate, the approach to equilibration would be
accelerated ($\lambda_1<0$) if the particles traveling in the prolate ($z$)
direction show a higher rate of collinear splitting, since such splitting is
an equilibrating process.  The rate of collinear splitting depends on the
efficiency of transverse momentum diffusion.  But the proto-plasma instability
caused by a prolate distribution is automatically the right one to enhance
such transverse momentum diffusion for particles moving along the $z$ axis.%
\footnote{
    It also slows down transverse momentum diffusion for particles in the
    ``equator'' of the prolate distribution, slowing their approach to
    equilibrium.  The angle averaged rate of splitting remains constant at
    this order.  But $\lambda_1$ does not depend
    on this angle averaged rate; it is dominated by what happens along
    the axis of prolateness (or oblateness).
    Therefore we can get the right sign by paying attention only to what
    happens to particles along the $z$ axis.
    }
Therefore the contribution of collinear splitting processes in $\C_{1;\M_1}$
should contribute negatively to $\lambda_1$ and give the first hints
of the effects of plasma instabilities.  

The fractional change in scattering efficiency due to $f_1$ grows at small
momentum exchange as $1/q^2$.  This behavior is expected; for weakly
anisotropic plasmas only the smallest $q$'s show plasma instabilities, which
appear in perturbation theory to give an infinite scattering rate.  Since
$\sigma_{ij}$ is treated as formally infinitesimal, there is no finite
momentum $q$ which becomes unstable, but the restoring effect of the plasma is
changed more and more for softer and softer magnetic $q$.
This leads to an IR log divergence in the total momentum transfer
rate $\int d^2 q_\perp q_\perp^2 C(q_\perp)$, see
Appendix \ref{subsec:collin}.  Therefore the change to the rate of collinear
splitting is log divergent when computed at leading perturbative order.

This means that our result for $\lambda_1$
actually includes a (negative in sign) logarithmically divergent
contribution, at least using the perturbative calculational tools we employ
here.  The log is $\ln(\mD/\epsilon)$, with $\epsilon$ an
artificially imposed minimum momentum transfer, implemented by
modifying $q^2 \rightarrow q^2+\epsilon$ in the denominator
for transverse gauge boson exchange when computing this process.

Physically, there really will be a limit on the infrared end of momentum
transfer.  In QCD we expect $\epsilon \sim g^2 T$ the magnetic screening
scale.  This is where the perturbative treatment of plasma corrections to
gauge field propagation breaks down.  Unfortunately we cannot compute the
exact form of this cutoff (the constant under the log,
$\ln(\mD/g^2 T) + k$) because this momentum region is strongly coupled.
Similarly, we expect that in QED the perturbative treatment of screening also
breaks down for $q \sim e^4 T$, where the physical distance of particle
propagation involved is of order the large-angle scattering length and the
electron propagators cease to behave like Eikonal propagators (as assumed in
the hard-loop computation of self-energies).  It might be
possible to compute the constant under the log,
$\ln(\mD/e^4 T) + k$ , but we have not done so.

As a result, we have not actually been able to compute the complete
finite-coupling value of $\lambda_1$.  Rather, we have guessed what the cutoff
$\epsilon$ on transverse momentum should be; we set $\epsilon=g^2 T/2$ in
QCD and $\epsilon = e^4 T/10$ in QED.  This leaves an uncertainty in our
results, set by the coefficient on the $\ln(\mD/\epsilon)$ term arising from
$C_{1;\M_1}$ from collinear splitting processes.  Fortunately, it turns out
that this contribution is numerically tiny.  If the constant under the log
shifts by 1 (the correct cutoff is $g^2 T/5.4$ rather than $g^2 T/2$)
then our result for $\lambda_1$ changes by less than $0.003$ in 3-flavor QCD
and less than $0.0003$ in pure-glue QCD or QED.

The extreme smallness of this effect arises as the product of several small
things. First, collinear splittings are not that important in driving
thermalization.  Second, the splitting rate is reduced in some directions, and
there is some angular averaging which reduces the total importance of the
shift in the splitting rate.  Third, the change to the splitting rate in any
specific direction also turns out to be numerically small.  This is another
indication that in practice the physical importance of plasma instabilities
turns out not to be very large.

We end the discussion by commenting about the range of validity of our
calculation.  In Figures \ref{fig:taupi}, \ref{fig:lambda1} we have plotted
our results out to $\mD/T = 3$, which corresponds to quite a large coupling
$\alphas = .48$ in 3-flavor and $\alphas=.72$ in pure-glue QCD.  The
calculation certainly cannot be believed at such couplings; probably it
becomes inadequate beyond $\mD/T=1$ (see \cite{CaronMoore} for a
next-to-leading order calculation of a similar transport coefficient).  The
scaled results for $\tau_\Pi$ and $\lambda_1$ are weakly dependent on details
of the theory, as shown by the almost identical results for $\lambda \phi^4$
theory and QCD at weak and relatively strong coupling.  But they rely in an
essential way on the validity of kinetic
theory.  There will be $\O(\alphas)$ corrections which cannot be incorporated
in kinetic theory, which we generically expect to change the shape of the
curves and which we do not know how to compute.  Therefore the flatness of the
curves in the figures can only be taken seriously at small $\mD/T$ (we would
guess below $\mD/T = 1.5$).

\medskip

{\bf Acknowledgements}

\medskip

We would like to thank Makoto Natsuume for suggesting we pursue this
calculation.  GM would like to thank the department of theoretical physics at
Bielefeld University for hospitality while this work was completed, and
the Alexander von Humboldt Foundation for its support through a F.\ W.\ Bessel
prize.  This work was supported in part by the Natural Sciences and
Engineering Research Council of Canada.

\appendix

\section{Matrix elements at nonzero $\sigma_{\mu\nu}$}
\label{sec:appendix}

A particle of momentum $P$ scattering from a particle of momentum $K$ via
gauge boson exchange with a soft exchange momentum $Q$, $|Q^2| \ll |P\cdot K|$
does so with a leading order matrix element (suppressing group factors)
\be
\M = 2P^\mu G_{\mu\nu} 2K^\mu \, , \qquad
G^{-1}_{\mu \nu} = Q^2 g_{\mu\nu} - Q_\mu Q_\nu - \Pi_{\mu\nu}[f]
+ \mbox{(Gauge fix)} \,,
\label{eq:Ginv}
\ee
with $G,\Pi$ understood as the retarded propagator and self-energy.
What is relevant here is that $\Pi_{\mu\nu}$ explicitly depends on the medium
through its distribution function $f$.  Write it as
$\Pi_{\mu\nu}[f] = \Pi_{\mu\nu,\rm eq} + \delta \Pi_{\mu\nu}[f_1]$
plus terms of higher order.  Then the squared matrix element becomes
\bea
|\M|^2 & = & \M_0 \M_0^* + \M_0 \M_1^* + \M_0^* \M_1 
   + \O(\lambda^2) \, , \nonumber \\
\M_0 & = & 2P^\mu G_{\mu\nu} 2K^\nu \, , \nonumber \\
\M_1 & = & 2 P^\mu G_{\mu\alpha} \, \delta \Pi^{\alpha \beta}[f_1] \,
G_{\beta \nu} 2K^\nu \,,
\label{eq:M_first}
\eea
where to simplify notation $G$ now means the equilibrium propagator.
Since $\Pi$ is suppressed relative to $G^{-1}$ unless $Q^2 \sim g^2 T^2$, we
can freely treat $Q^2$ as small in what follows, systematically expanding
whenever possible in $p,k \gg q,q^0$.  Similar expressions are also needed for
fermionic exchange processes and the fermionic self-energy.

Our goal in this appendix is to evaluate \Eq{eq:C1M1}.  Clearly as a first
step we need to evaluate $\delta \Pi^{\alpha\beta}$ and its fermionic
equivalent; then we need to use this to evaluate
$(\M_0 \M_1^* + {\rm h.c.})$ and perform the momentum integrations.
In addition, the collinear splitting rate is sensitive to $\delta \Pi$ because
it depends on the rate of soft momentum exchange; so we will have to revisit
the rate of collinear splittings as well.

\subsection{Bosonic self-energy}

With the sign convention established in \Eq{eq:Ginv},
for soft 4-momentum $Q=(q^0,\q)$ the leading order (retarded, hard-loop)
self-energy is \cite{MrowThoma}
\be
\Pi^{\mu\nu}(Q) = \sum_R g^2 T_R \int \frac{d^3 \p}{(2\pi)^3}
\frac{\partial f(p)}{\partial p^k} \left[ v^\mu g^{k\nu}
- \frac{v^\mu v^\nu q^k}{\v\cdot \q - q^0 -i\epsilon} \right]
\ee
where $\v=\p/p$ and $p=|\p|$ as usual.  The sum is over species, spin
and particle/antiparticle but not color.  Setting $f = f_0$ and using
\be
\sum_R g^2 T_R \int \frac{d^3 p}{(2\pi)^3} (-df_0/dp) = \mD^2=2m_g^2
\ee
recovers the usual HTL self-energies: in strict Coulomb gauge, which we use
henceforth,
\bea
G_{00} = \frac{1}{-q^2 - \Pi_{00}(Q)}
\equiv \frac{q^2-\omega^2}{q^2} G_L  \, ,&\quad&
 \Pi_{00}(Q) = m_g^2 \left( 2 - \frac{\omega}{q}\thelog \right) \,, \\
G_{ij} = \frac{\delta_{ij} - \hat{q}_i \hat{q}_j}{q^2-\omega^2-\Pi_T(Q)}
\equiv (\delta_{ij} - \hat q_i \hat q_j) G_T
\, , &\quad&
\Pi_T(Q) = -m_g^2 \left( \frac{\omega^2}{q^2} +
     \frac{\omega(q^2-\omega^2)}{2q^3} \thelog \right) \,. \nonumber
\eea
(Throughout the log has a $\mp i\pi$, with $-$ in retarded propagators $G,\Pi$
and $+$ in advanced propagators $G^*,\Pi^*$.)
Now we want to compute $\delta \Pi(Q)$ using
\be
\label{eq:deltaf}
f_1(p) = \frac{\beta\sigma_{ij}}{2} \symindex{v}{v}{i}{j} \chi(p)
\ee
where $\chi(p) = \beta^2 p^2 \tchi(p)$.  Then
\be
\frac{\partial f_1}{\partial p^k} = \frac{\beta \sigma_{ij}}{2} \left(
 \frac{v_i \delta_{jk} + v_j \delta_{ik} -2v_i v_j v_k}{p} \chi(p)
 + \symindex{v}{v}{i}{j} v_k \chi'(p) \right)\,.
\ee

The integration separates into an angular and a radial part.
Integrating the $\chi'$ radial term by parts gives
\bea
\delta \Pi_{\mu\nu}(Q) &=& \beta \left( \sum_R \frac{g^2 T_R}{2\pi^2}
\int p dp \chi(p) \right) \times A_{\mu\nu} 
\equiv  \beta \delta m_g^2 A_{\mu\nu} \, , \nonumber \\
A_{\mu\nu} & = & \frac{\sigma_{ij}}{2} \int d\Omega_\v
     \left( v_i \delta_{jk} + v_{j} \delta_{ik}
 - 4 v_i v_j v_k + \frac{2}{3} \delta_{ij} v_k \right)
 \left( v^\mu g^{k\nu} - \frac{v^\mu v^\nu q^k}{v_l q_l-q^0} \right)\,.
\eea
This depends on $Q$ only through $\hat{\q}$ and $q^0/q$; henceforth we rescale
$\q$ to be a unit vector, and $q^0 = \eta\equiv q^0/|\q|$.

First let us find $A_{00}$:
\be
A_{00} = - \frac{\sigma_{ij}}{2} \int d\Omega_\v 
      \frac{v_i q_j + v_j q_i-4 v_i v_j \v\cdot \q 
        + 2\delta_{ij} \v\cdot \q/3}{\v\cdot\q - q^0}
\equiv \frac{\sigma_{ij}}{2} \symindex{q}{q}{i}{j} A
\label{eq:A00}
\ee
(since this is the only possible tensorial structure).
To find $A$, contract the integral with $q_i q_j$ and define $x=\q\cdot \v$:
\be
A = \frac{3}{4} \int_{-1}^1 dx \frac{4x^3 - 8x/3}{x-\eta}
=  2(3\eta^2-1) - (3\eta^3 - 2\eta) \thelog  \,.
\label{eq:A}
\ee

Next consider $A^{0k}$:  in practice we will only need
\bea
(\delta_{kl} - q_k q_l) A^{0l} & = & \frac{\sigma_{ij}}{2}\int d\Omega_\v 
   \frac{ (-v_k+q_k q\cdot v) ( v_i q_j + v_j q_i - 4 v_i v_j \v\cdot \q
               + 2 \delta_{ij} \v\cdot \q/3 )}{\v\cdot\q - q^0}
\nonumber \\
& = & (\sigma_{ij}/2) B \left[ q_i \delta_{jk}+q_j \delta_{ik} 
             - 2 q_i q_j q_k \right] \,.
\eea
The coefficient is found by contracting with $q_i \delta_{jk}$:
\be
B = \frac{11\eta - 12 \eta^3}{6} + \frac{(1-\eta^2)(1-4\eta^2)}{4}
\thelog  \, .
\label{eq:B}
\ee

Finally we need $A_{lm}$.  In practice we need it only contracted against
transverse projectors:
\be
A_{lm} = \frac{\sigma_{ij}}{2} \int d\Omega_\v 
  \left( v_i \delta_{jk}+v_j \delta_{ik} -4v_i v_j v_k 
  + \frac{2}{3} \delta_{ij} v_k \right)
  \left( v_l g_{mk} - \frac{v_l v_m q_k}{\v\cdot\q - q^0} \right)
\ee
must be of form (defining $\hat\delta_{lm} \equiv \delta_{lm}-q_l q_m$)
\be
\hat\delta_{lr} A_{rs} \hat\delta_{ms}
 = \frac{\sigma_{ij}}{2} \left(\!
C_1 \hat\delta_{lm} \; \symindex{q}{q}{i}{j}
+C_2 \left[ \hat\delta_{il} \hat\delta_{jm}
  + (i\leftrightarrow j) - \frac{2 \delta_{ij}}{3}
  \hat\delta_{lm} \right] \right) .
\label{eq:Alm}
\ee
Contracting both the quantity in parenthesis and the original integral
expression with two independent tensors, such as $q_i q_j \delta_{lm}$ and
$\delta_{il} \delta_{jm}$, determines the coefficients:
\bea
C_1 & = & \frac{(1-\eta^2)(15\eta^2-4)}{6} +
   \frac{\eta(1-\eta^2)(3-5\eta^2)}{4} \thelog \, ,
\nonumber \\
C_2 & = & \frac{(1-\eta^2)(2-3\eta^2)}{6} 
    -\frac{\eta(1-\eta^2)^2}{4} \thelog \,.
\label{eq:c}
\eea

\subsection{Fermionic self-energy}
\label{sec:fermself}

The fermionic self-energy correction is \cite{MrowThoma}
(convention $1/(\nott{Q}-\Sigma)$)
\be
\Sigma(Q) = -\frac{g^2 \cf}{2\pi^2} \int \frac{p^2 dp}{2p}
\int d\Omega_{\p} \left[ 2 f_g + f_q + f_{\bar q} \right]
\frac{\hat \p \cdot \gamma - \gamma^0}{\hat\p \cdot \q - q^0}
\label{sigmaQ}
\ee
Hence the equilibrium value is
\be
\Sigma_{\rm eq}(Q) = \frac{g^2 \cf T^2}{16 q} \left(
\gamma_i \hat{q}_i \left[ -2 + \eta \thelog \right] 
- \gamma^0 \thelog \right) \,.
\ee
Taking $f_1$ from \Eq{eq:deltaf}, the correction term is
\bea
\dsig &=& \beta \left( \frac{g^2 \cf}{2\pi^2 q} 
      \int pdp (\chi_g(p)+\chi_q(p)) \right)
      \frac{\sigma_{ij}}{2} \int d\Omega_\v 
               \left( v_i v_j - \frac{\delta_{ij}}{3} \right)
               \frac{-v_k \gamma_k + \gamma^0}{\v\cdot\hat{q} - \eta}
\nonumber \\
& \equiv &  \beta \left( \frac{g^2 \cf}{2\pi^2 q} 
      \int pdp (\chi_g(p)+\chi_q(p)) \right) A 
\equiv \beta \frac{\delta m_f^2}{2q} A \,.
\eea
We have $A = A_0 \gamma^0 + A_k \gamma_k$.

Start with $A^0$:
\bea
A^0 &=& -\frac{\sigma_{ij}}{2} \int d\Omega_\v \frac{v_i v_j - \delta_{ij}/3}
       {\v\cdot\hat\q-\eta} \nonumber \\
&=& -\frac{\sigma_{ij}}{2} \; \symindex{q}{q}{i}{j}
       \left( \frac{3}{4} \int_{-1}^1 dx 
       \frac{x^2-1/3}{x-\eta} \right) \nonumber \\
& = & \frac{\sigma_{ij}}{2} \; \symindex{q}{q}{i}{j}
       \left( \frac{-3 \eta}{2} + \frac{3\eta^2-1}{4} \thelog \right) \,.
\eea
Similarly
\bea
A_k & = & -\frac{\sigma_{ij}}{2} 
        \int d\Omega_\v v_k \frac{v_i v_j-\delta_{ij}/3}
       {\v\cdot\hat\q - \eta} \nonumber \\
& = & \frac{\sigma_{ij}}{2} \left( 
       \kappa_1 q_k  \; \symindex{q}{q}{i}{j}
       +\kappa_2 \left[ q_i \delta_{jk} + q_j \delta_{ik}
                        -\frac{2}{3} \delta_{ij} q_k \right] \right)\,.
\eea
Determine the coefficients by contracting with $q_i q_j q_k$
and with $q_i \delta_{jk}$:
\be
\frac{2\kappa_1 +4\kappa_2}{3} = \frac{1}{6} \int_{-1}^{1} dx
\frac{x-3x^3}{x-\eta} \, , \qquad
\frac{2\kappa_1 + 10\kappa_2}{3} = \frac{1}{3} \int_{-1}^{1} dx
\frac{-x}{x-\eta} \,.
\ee
Therefore
\bea
\kappa_1 & = & \frac{4-15\eta^2}{6} 
               + \frac{-3\eta+5\eta^3}{4} \thelog \, , \nonumber \\
\kappa_2 & = &  \frac{-2+3\eta^2}{6} + \frac{\eta-\eta^3}{4}
                \thelog \,.
\eea   
Replacing $\gamma^\mu \rightarrow Q^\mu$ in \Eq{sigmaQ} gives an angular
averaged integral and so $Q_\mu \delta \Sigma^\mu = 0$, or
$\eta A^0 - (\kappa_1 + 2\kappa_2) = 0$, which is satisfied.  This is a fast
way to see that the correction to the hard propagation velocity $m_\infty^2$
is isotropic.

\subsection{Bosonic $2\leftrightarrow 2$ contribution to $C_{1;\M_1}$}
\label{sec:C1m1bos}

We work in the plasma rest frame and systematically approximate that the
incoming particle energies $p,k$ are much larger than the transfer momentum
$q$ or frequency $|q^0|\leq q$.  Using the integration variable
parametrization of \cite{AMY1}, the contribution
to $\Pi_{ij,\rm 2\; order}$
is
\bea
\Pi_{ij,{\rm 2\; order}} & \supset &
 \frac{A_{ab}}{2^8 \pi^5} \int_0 dp \int_0 dk \int_0 q dq
\int_{-1}^{1} d\eta \int_0^{2\pi} \frac{d\phi}{2\pi} \:
f_0(p)[1{\pm}f_0(p)] f_0(k) [1{\pm} f_0(k)]
\nonumber \\ && \times
T \bchi_{ij}(p)
\frac{\sigma_{rs}}{2} \left( \bchi_{rs}(\p) + \bchi_{rs}(\k)
  -\bchi_{rs}(\p') -\bchi_{rs}(\k') \vphantom{\Big|} \right)
\Big( \M_0^* \M_1 + \M_0 \M_1^* \Big).
\qquad \;
\label{eq:C1M1a}
\eea
Here $\M_0,\M_1$ are to be normalized as in \Eq{eq:M_first};
we have absorbed all color factors into $A_{ab}$ which in SU($\nc$) gauge
theory with $\nf$ fermions is
$16 \df^2 \nf^2 \cf^2/\da$ for fermion-fermion scattering,
$16 \df \nf \cf \ca$ for fermion-boson scattering and
$4 \da \ca^2$ for boson-boson scattering.  Symmetry between
$p,p',k,k'$ allows us to replace
\be
\bchi_{ij}(p) \rightarrow \frac{1}{4}
    \Big( \bchi_{ij}(\p)+\bchi_{ij}(\k)
         -\bchi_{ij}(\p')-\bchi_{ij}(\k') \Big)
\ee
and small $q$ approximations allow \cite{AMY1}
\be
    \Big( \bchi_{ij}(\p)+\bchi_{ij}(\k)
         -\bchi_{ij}(\p')-\bchi_{ij}(\k') \Big)
\simeq -q\beta^3 \Big(
     2\hat{\p}_{\langle i} \hat{\q}_{j\rangle} p \bchi(p)
     + \eta \hat{\p}_{\langle i} \hat{\p}_{j\rangle} p^2 \bchi'(p)
     - (p\rightarrow k) \Big)
\ee
and similarly for the $\bchi_{lm}$ term.

All angles are determined by the $\eta,\phi$ variables; in particular
$x_{pq}=\eta=x_{kq}$ and $x_{pk} = \eta^2 + (1{-}\eta^2) \cos\phi$.
Therefore, extracting a factor of $pk$ from $\M_0$ and $\M_1$,
$\tilde\M_0 \equiv \M_0/pk$,
the integrals over the magnitudes $p,k$ factorize from the integrals over
$q,\omega,\phi$.  Defining the integrals
\be
\left. \begin{array}{c} K_0 \\ K_1 \\ K_2 \\ K_3 \\ K_4 \\ K_5 \\
       \end{array} \right\} = \int_0^\infty dp p^2 (-f_0'(p)) \times 
     \left\{ \begin{array}{l} 1 \\ 4p^2 \bchi^2 \\ 4p^3 \bchi \bchi'
         \\ p^4 (\bchi')^2 \\ 2p\bchi \\ p^2 \bchi' \\ \end{array}
       \right.
\ee
we need, for the double fermion term for instance,
\bea
\frac{\beta^5 A_{ff}}{2^{10} \pi^5}&& \int_0 q^3 dq \int_{-1}^1 d\eta 
     \int \frac{d\phi}{2\pi}
\Big( \tilde\M_0 \tilde\M_1^* + \tilde\M_0^* \tilde\M_1 \Big)
\frac{\sigma_{lm}}{2}
\nonumber \\ && \times 
\Big(
     2K_0 K_1       \symindex{\hat\p}{\hat\q}{i}{j}
                    \symindex{\hat\p}{\hat\q}{l}{m}
    +2K_0 K_2\eta   \symindex{\hat\p}{\hat\q}{i}{j}
                    \symindex{\hat\p}{\hat\p}{l}{m}
    +2K_0 K_3\eta^2 \symindex{\hat\p}{\hat\p}{i}{j}
                    \symindex{\hat\p}{\hat\p}{l}{m}
\nonumber \\ && \qquad \;
    -2K_4^2         \symindex{\hat\p}{\hat\q}{i}{j}
                    \symindex{\hat\k}{\hat\q}{l}{m}
    -4K_4 K_5\eta   \symindex{\hat\p}{\hat\p}{i}{j}
                    \symindex{\hat\k}{\hat\q}{l}{m}
    -2K_5^2\eta^2   \symindex{\hat\p}{\hat\p}{i}{j}
                    \symindex{\hat\k}{\hat\k}{l}{m}
\Big)\,,
\label{eq:horrible}
\eea
where we used $p\leftrightarrow k$ symmetry to simplify some terms.
The matrix element squared is
\bea
\label{eq:M0M1}
\tilde\M^*_0 \tilde\M_1 & = & 16
    \left( G_{00}^* + (1{-}\eta^2) \cos\phi G_T^* \right)
\times  \beta \delta m_g^2 \frac{\sigma_{rs}}{2} 
\\ && \times 
     \bigg( G_{00}^2 A \dhat{q}{q}
      + G_{00} G_T B \left[ \dhat{\p}{\q} + \dhat{\k}{\q}
                           -2\eta  \dhat{\q}{\q} \right]
\nonumber \\ && \quad\;
      + G_T^2 \left[ C_1 (1{-}\eta^2) \cos\phi \dhat{\q}{\q}
               +2C_2 ( \dhat{\p}{\k} - \eta \dhat{\k}{\q} - \eta \dhat{\p}{\q}
                   + \eta^2 \dhat{\q}{\q} ) \right] \bigg).
\nonumber
\eea
The integral $\int_0^{2\pi} \frac{d\phi}{2\pi}$ can always be done analytically
by replacing $\cos^{(0,1,2,3,4)}\phi = (1,0,\frac{1}{2},0,\frac{3}{8})$.
Using repeatedly \Eq{eq:niftyangle} the evaluation of \Eq{eq:horrible} is now
straightforward, if lengthy.

One potential pitfall in performing the $q,\eta$ integrals in \Eq{eq:horrible}
is the possibility of an infrared small $q$ divergence.
This can come about because $G_T(q,\eta)$ behaves, for $q<\mD$ and
$\eta<q^2/\mD^2$, like $G_T \sim 1/q^2$.  The integration region over which
this behavior applies is $q^5 dq$ but the $G_T^* G_T^2$ term in
\Eq{eq:M0M1} is $1/q^6$ so there is a potential log divergence.
To determine whether this divergence occurs it is sufficient to approximate
$\eta = 0$ in the integrands, other than in $G_T$.  In this limit
$C_1 = -2C_2$.  Only the $K_0 K_1$ and $K_4^2$ terms are zero-order in $\eta$
so only they need be computed; the relevant global angular averages are
\bea
&& K_0 K_1 (\ldots) \cos\phi \int d\Omega_{\rm global}
\symindex{\hat\p}{\hat\q}{i}{j}\symindex{\hat\p}{\hat\q}{l}{m}
\Big( \dhat{\p}{\k} - \cos\phi \dhat{\q}{\q} \Big) \,,
\nonumber \\
&& K_4^2 (\ldots)  \cos\phi \int d\Omega_{\rm global}
\symindex{\hat\p}{\hat\q}{i}{j}\symindex{\hat\k}{\hat\q}{l}{m}
\Big( \dhat{\p}{\k} - \cos\phi \dhat{\q}{\q} \Big) \,.
\eea
Applying \Eq{eq:niftyangle} setting $x_{pq}=0=x_{kq}$ and
$x_{pk}=\cos\phi$ and averaging over $\phi$,
one finds that each term happens to vanish, so
the potential IR divergence does not occur.

\subsection{Fermionic $2\leftrightarrow 2$ contribution to $C_{1;\M_1}$}

The infrared region of virtual fermion exchange is also important at leading
order for transport \cite{AMY1}.  The contribution is still described by
\Eq{eq:C1M1a} but with $A=32\nf \cf^2 \df$ each for pair annihilation and
Compton scattering.  Since the matrix element is less infrared singular, we
can approximate $\bchi_{rs}(\p)= \bchi_{rs}(\p')$ and similarly for $k$.  But
if $\bchi(p)$ represents a fermion, then $\bchi(k),\bchi(k')$ represent a
quark and gluon for annihilation, but a gluon and quark for Compton
scattering.  Therefore, summing over the processes, the $p,k$ cross-terms
cancel and we may approximate
\be
\bar\chi_{ij}(p)  \left( \bchi_{rs}(\p) + \bchi_{rs}(\k)
  -\bchi_{rs}(\p') -\bchi_{rs}(\k') \vphantom{\Big|} \right)
= \frac{1}{2} \Big( \bar\chi_{ij,q}(p)-\bar\chi_{ij,g}(p) \Big)
 \Big( \bar\chi_{lm,q}(p)-\bar\chi_{lm,g}(p) \Big)
\ee
where the subscripts $q,g$ indicate if the species is a quark or a gluon.
This simplifies matters considerably; pulling a factor $pk$ out of
$\M^2$, the $p,k$ integrals we need are
\be
\beta^5 \int dp p^5 f_{0,f}(p) [1{+}f_{0,b}(p)] (\bchi_q-\bchi_g)^2 \;
\int dk k f_{0,f}(k) [1{+}f_{0,b}(k)]
\ee
which multiply the $q,\eta$ integral
\be
\int q dq \int_{-1}^1 d\eta \int \frac{d\phi}{2\pi} \frac{\sigma_{lm}}{2}
\hat{p}_{\langle i} \hat{p}_{j\rangle}
\hat{p}_{\langle l} \hat{p}_{m\rangle}
\Big( \tilde{\M}_0^* \tilde{\M}_1 + {\rm h.c.} \Big)
\ee
with
\bea
\tilde\M_0^* \tilde\M_1 & = & \frac{1}{(\tilde Q^2)^2 (\tilde Q^*)^2}
\Tr \hat{\nott{p}} \nott{\tilde Q} \delta \nott\Sigma
\nott{\tilde Q} \hat{\nott{k}} \nott{\tilde Q}^*
\eea
with $\tilde Q^\mu \equiv Q^\mu - \Sigma^\mu_{\rm eq}$ and
$\delta \Sigma^\mu$ as given in Appendix \ref{sec:fermself}.
The trace and global angular average are straightforward but tedious.

\subsection{Collinear $1\leftrightarrow 2$ contribution to $C_{1;\M_1}$}
\label{subsec:collin}

According to \cite{AMY6}, the rate at which a particle in the thermal medium
splits into two is given by
\bea
\C_{1\leftrightarrow 2}[f(p)] & = & \frac{(2\pi)^3}{2p^2} \sum_{bc}
\int_0^\infty dp' dk' \delta(p-p'-k') \gamma^a_{bc}(p,p',k')
\nonumber \\ && \times
\Big( f(\p) [1{\pm}f(\k')][1{\pm}f(\p')]
       - [1{\pm}f(\p)] f(\k') f(\p') \Big)
\label{eq:C12}
\eea
where $\p,\k'\p'$ are collinear at leading order, that is, $\k'=k\hat\p$.
We saw how this term gives rise to contributions to $\C_{11}$.  It also
contributes to $\C_{1;\M_1}$ because the splitting rate $\gamma^a_{bc}$ is
sensitive to the details of the plasma, and can be expanded as
\be
\gamma^a_{bc} = \gamma^a_{bc,0} + f_1 \gamma^a_{bc,1} + \ldots
\ee
We need to evaluate $\gamma^a_{bc,1}$; it will then contribute to
$\C_{1;\M_1}$ through \Eq{eq:C12} with the population functions (second line)
replaced by 
\be
\Rightarrow \quad f(k')f(p')[1{\pm}f(p)]\;
\beta^5
\; p^2\bchi(p) \Big( p^2\bchi(p)-k'{}^2 \bchi(k)-p'{}^2\bchi(p') \Big)\;
\hat{p}_{\langle i}\hat{p}_{j\rangle}
\hat{p}_{\langle l}\hat{p}_{m\rangle} \;\frac{\sigma_{lm}}{2} \,.
\ee

Besides overall coefficients tabulated in \cite{AMY6}, $\gamma^a_{bc}$ is
proportional to the integral over the solution to an integral equation:
\bea
\gamma^a_{bc} & \propto &
\int d^2\h 2\h \cdot \F \, , 
\nonumber \\
2\h & = & (i\delta E) \F(\h) + \int \frac{d^2 \q_\perp}{(2\pi)^2}
  C(\q_\perp) \Big\{ 
    (C_{\rm s}-{\textstyle \frac{\ca}{2}}) [\F(\h) - \F(\h-k'\q_\perp)]
\nonumber \\ && \hspace{3.4cm}
+{\textstyle \frac{\ca}{2}} [ \F(\h) - \F(\h-p'\q_\perp)]
+{\textstyle \frac{\ca}{2}} [ \F(\h) - \F(\h+p\q_\perp)] \Big\}\,. 
\qquad \quad
\label{eq:F}
\eea
Here $\h$ is a vector in the 2-component space transverse to $\p$,
$\delta E$ is medium dependent but in a way which is insensitive
to $f_1$ (see footnote \ref{footnote:mass}); however $C(q_\perp)$, which
represents the differential rate to scatter with transverse momentum transfer
$\q_\perp$, is sensitive.  Explicitly,
\be
C(\q_\perp) = \int \frac{dq_z}{2\pi} G^>_{++}(q^0=q_z,q_\perp)
\ee
with $G^>_{++}$ the gauge boson Wightman function, equal to
$T/\omega$ times the discontinuity in the retarded function.  In Coulomb gauge
this is
\be
G^>_{++}(Q) = \frac{2T}{q^0} \:{\rm Disc}\: 
   \left( G_{00} + G_{T,zz} \right) \,.
\ee
Here the retarded Green functions $G_{00}$, $G_T$ include the first order
corrections, that is, $G_T = G_{T,0} + G_{T,0}\delta \Pi_T G_{T,0}$.
According to \cite{Simon}, analyticity properties allow for the simple
evaluation of this integral:
\be
C(\q_\perp) = T \Big( G_{T,zz}(0,0,\q_\perp) + G_{00}(0,0,\q_\perp) \Big)\,.
\ee
In equilibrium this reproduces the sum rule of Aurenche, Gelis, and Zaraket
\cite{AGZ},
\be
C(\q_\perp) = T \left( \frac{1}{\q_\perp^2} - \frac{1}{\q_\perp^2+\mD^2}
    \right) \,.
\ee
For our application the first order shift is
\be
C_1(\q_\perp) = T \left( \frac{\delta \Pi_{T,zz}(\eta=0)}{\q_\perp^4}
  + \frac{\delta \Pi_{00}(\eta=0)}{(q_\perp^2+\mD^2)^2} \right)
\ee
Using \Eq{eq:A00} , \Eq{eq:A}, \Eq{eq:Alm}, and \Eq{eq:c},
\bea
\delta \Pi_{00} & = & \beta \delta m_g^2 \frac{\sigma_{rs}}{2}
  \dhat{q}{q} (-2) \,, \nonumber \\
\delta \Pi_{T,zz} & = & \beta \delta m_g^2 \frac{\sigma_{rs}}{2}
  \left( -\frac{2}{3} \dhat{q}{q} + \frac{2}{3} \dhat{p}{p} \right)
\eea
where we used that the $z$ direction means the $\hat{p}$ direction.

If $\hat{q} = \hat{x}\cos\phi + \hat{y} \sin\phi$ then
\bea
\dhat{q}{q} & = & \hat q_r \hat q_s-\frac{\delta_{rs}}{3}
= -\frac{1}{3} \delta_{rz} \delta_{sz} 
+ \frac{1}{6} ( \delta_{rx}\delta_{sx} + \delta_{ry}\delta_{sy} )
\nonumber \\ && \qquad \qquad
+ \frac{1}{2} \Big( (\delta_{rx} \delta_{sx} - \delta_{ry} \delta_{sy})
           \cos2\phi + 
           (\delta_{rx}\delta_{sy}+\delta_{sx}\delta_{ry})\sin2\phi ) \Big)
\nonumber \\
& = & -\frac{1}{2} \dhat{p}{p} + \O(\cos^2\phi,\sin^2\phi) \,.
\eea
When expanding \Eq{eq:F} to linear order in $C_1$ the $\phi$ dependent terms
will yield $\phi$ dependence in $\F$ which cancels on angular $\h$
integration; therefore these terms may be dropped and
$\dhat{q}{q}$ replaced with $-\dhat{p}{p}/2$.  Hence
\be
\delta \Pi_{00} = \beta \delta m_g^2 \frac{\sigma_{rs}}{2}
\dhat{p}{p} = \delta \Pi_{T,zz}
\ee

To evaluate the shift induced by the correction we have found to $\C_1$, we
should expand \Eq{eq:F} linearly in the correction to $C(\q)$:  schematically
(recycling the inner product notation for functions over $\h$ with
$\int d^2 \h$ as inner product)
\bea
|2\h \rangle & = & ( i \delta E + C_0 + C_1 ) | \F \rangle \, ,
\nonumber \\
|\F\rangle & = & \left( \frac{1}{i\delta E + C_0} -
  \frac{1}{i\delta E + C_0} C_1 \frac{1}{i\delta E + C_0} 
   + \O(C_1^2) \right)
|2\h \rangle \,.
\eea
The tools for solving this integral equation are similar to those used in
solving the Boltzmann equation.  The integral we need
is $\langle 2\h \,|\, \F \rangle$.  With explicit formulae for everything, the
result of the analysis is {\em almost} straightforward.

There is one complication, however.  Plugging it all in,
\be
C_1(\q_\perp) = \left( \delta m_g^2 \dhat{p}{p} \frac{\sigma_{rs}}{2}
  \right)
\left( \frac{1}{\q_\perp^4} + \frac{1}{(\q_\perp^2+\mD^2)^2} \right)
\ee
has a $1/\q_\perp^4$ singularity at small $q_\perp$.  Together with the
integration measure $d^2 \q_\perp$ and the $\F$ differencing, which on angular
averaging behaves like 
$F(\h)-F(\h+a\q_\perp) \sim a^2 q_\perp^2 \nabla^2 F(\h)$, the rest of the
integration behaves like $q_\perp^3 dq_\perp$, resulting in a log IR
divergence.  The divergence is cut off at large momenta by the Debye scale,
where $\F$ starts to display more complicated behavior.  In the infrared the
calculation becomes unreliable at exchange momentum $\q_\perp\simeq g^2 T$
where the perturbative expansion breaks down.  We expect that in a nonabelian
gauge theory the divergence is cut off at this scale, but we are unable to
compute the IR end in detail.  In order to push forward with the calculation
we cut the integral off by replacing $1/q_\perp^4$ with
$1/(q_\perp^2+(\epsilon \mD)^2)^2$ in the denominator, which allows to extract
the coefficient and constant under the log.   However, the contribution to
$\lambda_1$ arising from collinear contributions to $\C_{1;\M_1}$ is
numerically very small, and the coefficient of this log is still smaller,
never exceeding 0.003 for 3-flavor QCD and 0.0003 for pure-glue QCD.
Therefore in practice the uncertainty from resolving this logarithm is too
small to see in Figure \ref{fig:lambda1}.

\end{document}